\documentclass{bmcart}\usepackage[]{graphicx}\usepackage[]{color}
\makeatletter
\def\maxwidth{ %
  \ifdim\Gin@nat@width>\linewidth
    \linewidth
  \else
    \Gin@nat@width
  \fi
}
\makeatother

\definecolor{fgcolor}{rgb}{0.345, 0.345, 0.345}

\usepackage{framed}
\makeatletter
 {\par\unskip\endMakeFramed%
 \at@end@of@kframe}
\makeatother

\definecolor{shadecolor}{rgb}{.97, .97, .97}
\definecolor{messagecolor}{rgb}{0, 0, 0}
\definecolor{warningcolor}{rgb}{1, 0, 1}
\definecolor{errorcolor}{rgb}{1, 0, 0}
\newenvironment{knitrout}{}{} 

\usepackage{alltt}

\usepackage{amsthm,amsmath}
\usepackage{multirow}
\usepackage{rotating}
\usepackage[utf8]{inputenc} 

\startlocaldefs
\endlocaldefs

\IfFileExists{upquote.sty}{\usepackage{upquote}}{}
\begin{document}

\begin{frontmatter}

\begin{fmbox}
\dochead{Research}


\title{Chatbots and messaging platforms in the classroom: an analysis from the teacher's perspective}

\author[
  addressref={aff1},
  corref={aff1},
  email={jmerelo@ugr.es}
]{\inits{JJM}\fnm{Juan J.} \snm{Merelo}}
\author[
  addressref={aff1},
  email={pacv@ugr.es}
]{\inits{PAC}\fnm{Pedro A.} \snm{Castillo}}
\author[
  addressref={aff2},
  email={amorag@ugr.es}
]{\inits{AMM}\fnm{Antonio M.} \snm{Mora}}
\author[
  addressref={aff1},
  email={fbarranco@ugr.es}
]{\inits{FB}\fnm{Francisco} \snm{Barranco}}
\author[
  addressref={aff3},
  email={n.h.abbas@leeds.ac.uk}
]{\inits{NA}\fnm{Noorhan} \snm{Abbas}}
\author[
  addressref={aff1},
  email={aguillen@ugr.es}
]{\inits{AG}\fnm{Alberto} \snm{Guillén}}
\author[
  addressref={aff4},
  email={o.tsivitanidou@cyens.org.cy}
]{\inits{OT}\fnm{Olia} \snm{Tsivitanidou}}

\address[id=aff1]{%
  \orgdiv{Department of Computer Architecture and Technology},
  \orgname{University of Granada},
  \city{Granada},
  \cny{Spain}
}
\address[id=aff2]{%
  \orgdiv{Department of Signal Theory, Telematics and Communications},
  \orgname{University of Granada},
  \city{Granada},
  \cny{Spain}
}
\address[id=aff3]{%
  \orgdiv{School of Computing},
  \orgname{University of Leeds},
  \city{Leeds},
  \cny{UK}
}
\address[id=aff4]{%
  \orgname{CYENS},
  \city{Nicosia},
  \cny{Cyprus}
}

\end{fmbox}


\begin{abstractbox}

\begin{abstract} 
Introducing new technologies such as messaging platforms, and the chatbots
attached to them, in higher education, is rapidly growing. This introduction
entails a careful consideration of the potential opportunities and/or challenges
of adopting these tools. Hence, a thorough examination of the teachers’
experiences in this discipline can shed light on the effective ways of
enhancing students’ learning and boosting their progress. In this contribution,
we have surveyed the opinions of tertiary education teachers based in
Spain (mainly) and Spanish-speaking countries. The focus of these surveys is to collect
teachers’ feedback about their opinions regarding the introduction of the messaging
platforms and chatbots in their classes, understand their needs and to gather
information about the various educational use cases where these tools are
valuable. In addition, an analysis of how and when teachers’ opinions towards
the use of these tools can vary across gender, experience, and their discipline
of specialisation is presented. The key findings of this study highlight the
factors that can contribute to the advancement of the adoption of messaging
platforms and chatbots in higher education institutions to achieve the desired
learning outcomes.
\end{abstract}



\end{abstractbox}
%

\end{frontmatter}


\section{Introduction}

The introduction of new technologies in the classroom, \textcolor{black}{to be successful, involves extra
teacher training, devising methods for enhancing student engagement with the new
technology, and indeed acquiring new skills by students and teachers alike}.
Some technologies are readily adopted, but others require longer time for
full adoption. In most cases, the magnitude of the uptake of the new technology
impinges on the collaboration of all the parties using it. \textcolor{black}{Hence, examining the
users’ opinions is an important first step in boosting this collaboration
and reaping the benefits of the technology}.  Recently, a surge in the use of
synchronous instant messaging tools in higher education such as WhatsApp,
Telegram and Facebook Messenger took place. Some messaging applications have
built-in chatbots to support a synchronous conversation between the different
parties \cite{differ.chat}, while other chatbots are developed as standalone
systems and can possibly be attached to a messaging application using an API
(Application Program Interface).

A \textit{chatbot} is a program, sometimes developed using Artificial
Intelligence techniques, able to communicate in a similar way as humans do
\cite{Gong2008}, using, in some cases, natural language. Indeed, most of these
applications would be, apparently, close to passing a classic Turing test
\cite{moor2003turing}, since they are able to answer almost any question
fluently, and even pose own questions in their conversations. The recent
advancement in this discipline has played an important role in many fields
especially in education and online tutoring
\cite{clarizia2018,Smutny2020}. These automatic systems facilitate the delivery
of personalized learning by adapting to students’ pace of learning and providing
customized online tutoring outside the classroom. Thus, chatbots can
significantly contribute to providing interactive learning experiences as well
as improving individual attention \cite{agarwal2020}.  In this sense, chatbot
technology offers a great opportunity for the improvement of tutoring systems
\cite{Daniel2016,agarwal2020}, as not all students are comfortable with
face-to-face tutoring with the instructor. In fact, in many cases, some students
experience stress because of their need to ask a question in front of the entire
class. This is why many students resort to silence in the classroom and tend to
contact their teachers later via email.  Consequently, this can lead to not only
a delay in obtaining answers to their questions but can also cause a significant
increase in their teachers’ workload, especially when factoring in the
disproportion between the class size and the number of teachers. Hence, the
chatbot technology has a potential to mitigate this problem by providing answers
to students’ questions and facilitating a dynamic and autonomous learning
experience \cite{Griol2014,kim2020}. In addition, using an automated system such
as a chatbot, can draw teachers’ attention to topics that students struggle to
comprehend or need further assistance in understanding.

Given the potential benefits that chatbots could bring to the classroom, the
focus of the
EDUBOTS KA2 European project \footnote{https://www.edubots.eu},
the sponsor of this study, is to
\textcolor{black}{explore best practices and innovative use of chatbots, and to create 
a learning community of educators in higher education institutions.
The majority of studies conducted so far have studied how these 
messaging applications and/or chatbots are used to deliver personalised
learning in classrooms that occurs anytime anywhere,
promote collaborative learning experiences, group discussions \cite{panah2020study}
and boost students' sense of belonging to their institutions
\cite{abbas2021onlinechat}. Hence, due to the paucity of research
that explores the use of these technologies from the teachers' perspectives,
this paper aims to investigate their opinions/needs and the
challenges/opportunities of adopting these technologies in classrooms
to paint a better picture of how they can positively contribute to enhancing
the learning process in higher education institutions.}

\textcolor{black}{In this study, data collection involved two phases:}
 \begin{itemize}
\item The first phase aimed to collect feedback from students about their actual
  information about students’ preferences of tools/applications for chatting and
  messaging, how they use them in educational context, who do they like to be
  with in the class messaging groups, and their expectations of the chatbots in
  terms of assisting them during their learning process. Therefore, two surveys
  for bachelor and master degree students at the University of Granada (Spain)
  were designed and answers from more than 250 students were collected. The key
  findings of the students' surveys reveal that they prefer to use messaging
  applications, such as Telegram or WhatsApp as they are familiar to
  them. Students also expressed interest in chatbots that can assist them with
  organizing their course schedules, help them access their assignment grades
  and facilitate searching for course's resources.
\item The results of the first phase were fed into the second phase of the
  study, which focused on teachers’ opinions about using chatbots and messaging
  applications in classes. This is what is going to be presented in this paper.
\end{itemize}

Initially, a single survey was developed for this purpose and responses from 300
higher education teachers were collected and analysed\textcolor{black}{; this will be explained more fully in the methodology section}. The responses' analysis
entailed the creation of a second survey to which 200 teachers' responses were
collected\textcolor{black}{, with the methodology and collective characteristics that will be explained below. The questions of the second survey focussed on discussing teachers'
needs, opinions and preferences of the development of future technology-enhanced
tools and its potential impact on educational institutions’ policies. In order to assess these, this paper investigated the following research questions}:

\begin{itemize}
\item RQ1 - Are teachers already using messaging apps in their classes?
\item RQ2 - Which chatbots' features would teachers find useful in their
  classes?
\item RQ3 - Which kind of interaction do teachers prefer with their students?
\item RQ4 - What kind of interaction media features do teachers value the most?
\end{itemize}

We will try to answer these questions through the analysis of questions and
answers in the two surveys. Eventually, what we want to obtain is a series of
recommendations to make a successful deployment of chatbot technologies (and, in
some cases, general instant messaging applications) in higher education\textcolor{black}{; from the response to RQ1 and RQ4 we will try to recommend specific technologies or applications to be used in the classroom; the response to RQ2 and RQ3 will help us recommend chatbots features or specific platforms; and finally, from RQ3 we will also try to find best practices in the adoption of messaging platforms and they accompanying chatbots.}

The remainder of the paper is organized as follows: \textcolor{black}{first, an overview of what
current research has found about the use of messaging applications, including
chatbots, in the classroom is described}. The methodology used in the surveys is presented in
Section \ref{sec:meth}, and the results of the surveys are presented next in
Section \ref{sec:res}. Finally, we discuss these results and conclude with a
series of recommendations in the section that closes the paper.


\section{State of the art}
\label{sec:soa}

The widespread and rapid adoption of free Mobile Instant Messaging
(MIM) tools/platforms such as WhatsApp, Telegram, WeChat and Facebook
Messenger stems from their simplicity, ease of use and multi-modality
(i.e. video, audio, text) \cite{tang2017mobile}. Using these tools in
higher education can facilitate the delivery of personalised learning
that occurs anytime anywhere, promote collaborative learning
experiences and group discussions \cite{panah2020study}.

WhatsApp is, at least in most Western countries, the most popular MIM platform
used by educators to \textcolor{black}{provide} assignments' feedback to students, support course
discussions, and provide learning resources in informal learning settings
\cite{panah2020study}.  Moreover, the use of WhatsApp in higher education could
enhance social presence \cite{tang2017mobile} and foster trust relationships
between educators and students embedded in the social learning process
\cite{gachago2015crossing}; however, this last paper also reflects the need for
learners to "take ownership of the tool" and the advantages of social learning
in general. At the same time, it also mentions different challenges, among which
the most important is the blurring of social and academic life; indeed, there are challenges when using MIM tools that occur due to the
blurring of boundaries between academic and private life. This can lead to
technostress \cite{gachago2015crossing}, difficulty in managing
responsibilities, especially among mature students, and lack of privacy
\cite{tang2017mobile}.  Students' dropout of the MIM groups, as they can leave
groups at any time, can hinder their learning and undermine educators' efforts
\cite{mwakapina2016whatsapp}.  In addition, there is a need to set rules and
norms for these MIM groups in order to maintain the safety of these online
communities for students \cite{abbas2021onlinechat}. However, these rules should
not affect students' ownership and control, since it is vital to advance in
their learning \cite{gachago2015crossing} process.  This is why examining the
role of MIM in higher education is still a challenge, and why the opinions of
the teaching community towards them have to be examined, as we do in this paper.

The use of MIM, although possibly valuable by itself, can be enhanced via the
use of chatbots, which, being conversational agents, usually dwell in systems
where synchronous conversations take place.  The use of conversational agents
(chatbots) in higher education is still at its infancy
\cite{yang2019opportunities}. Nevertheless, recent studies examining their
positive impact on students' academic performance \cite{perez2020rediscovering}
and engagement \cite{differ.chat,abbas2021onlinechat} have led to a growing
interest in using this technology in the (possibly virtual) classroom. Indeed,
the use of chatbots in collecting course feedback from students in higher
education improved students' response quality and boosted their enjoyment levels
\cite{abbas2021Surveys}.  According to \cite{Roblyer2010}, using either tools
such as mobile devices or teaching strategies based on gamification
\cite{Yildirim2017} can improve student motivation. In this sense, the authors
in \cite{Pimmer2019} \textcolor{black}{adopted a quasi-experimental, survey-based approach to
report the positive impact of using instant messaging tools in boosting
students' knowledge and mitigating their feelings of isolation}.

Several higher education chatbots' evaluation studies have been undertaken. For
instance, a recent evaluation review study presented by Smutny and Schreiberova
\cite{Smutny2020} examined 47 educational chatbots implemented in Facebook
Messenger with the focus to identify characteristics and quality metrics such as
language, subject matter and platform\textcolor{black}{, whereas the study undertaken by Pérez and collaborators
\cite{perez2020rediscovering} aimed to categorise educational chatbots, according
to their, purpose into service-oriented and teaching-oriented}. The first category
includes those that provide service support such as the chatbot Ask Holly
\cite{meetholly} and Dina \cite{santoso2018dinus}; both chatbots respond to
students' questions about enrolment and registration. Ask L. U.
\cite{LancasterUniversity_2019} answers students' frequently asked questions
about timetables, grades, tutors and societies. LISA \cite{dibitonto2018chatbot}
and Differ \cite{differ.chat} facilitate breaking the ice between new students
by introducing them to each other. Ranoliya et al. \cite{ranoliya2017chatbot}
proposed a generic chatbot for university students that is able to answer to
their frequently asked questions \textcolor{black}{and, for some time, the University of Granada had
\textcolor{black}{Elvira} as a chatbot in its main web page \cite{MOREO20129} to perform that duty.} Besides being able
to answer pre-established (frequent) questions, it did so from the website of
the University of Granada using an inset \textcolor{black}{persona, voiced by a real person, who
lip-synced the answers. The emphasis in the case of Elvira, however, was on the
authenticity of the synchronization more than in making the update of questions
answered simpler, or more interactive, or more open to the rest of the university staff, using actual messaging platforms. In
most cases, the embedded search engine was able to provide more up-to-date
and accurate answers than the ones provided by Elvira, which had to be updated
by hand (and not too often). It was eventually discontinued, and its technology was not adapted to other platforms, since it did not really provide an useful service. Its service was not tied to teaching anyway; it was more related to directory queries and university-wide administrative questions, so even if it is strictly an example of a chatbot in a university setting, it is not really a chatbot that can be adopted by teachers to affect the learning outcomes, something which is the focus of this paper}.

On the other hand, teaching-oriented chatbots are more sophisticated, as they
set personalised learning outcomes and monitor learning progress. For instance,
\cite{fernoagua2018intelligent} reported on ``eduAssistant'', a virtual teaching
assistant chatbot developed on the Telegram messaging platform. In this study,
the Telegram platform is chosen because it is easy to use, students are familiar
with its features, and it enables them to exchange messages in different formats
(text, audio and video) \cite{fernoagua2018intelligent}. In addition, Telegram
could operate on all devices and operating systems. The ``eduAssistant'' chatbot
acts as an automatic agent in teacher-content-student, facilitating real-time
feedback loops and providing a personalised learning experience relevant to the
students' acquired skills and knowledge. Using this chatbot, educators can
create interactive instances in their lectures where they pose questions to
their students and the chatbot assists those who need further help by giving
them more hints and reporting it to their educator's dashboard
\cite{fernoagua2018intelligent}. This can help educators locate those
students that need more attention and send them more educational resources
relevant to their academic attainment. \textcolor{black}{In addition, a recent study
has analysed how the use of chatbots positively affected the learning outcomes
of students in a Chinese class \cite{doi:10.1177/0735633120929622}}.

Despite its (arguably) successful implementation in different higher education
institutions, over all in the fringes of educational activity, not in the actual
student-teacher interaction, its implementation or deployment is not trivial.
Some authors, for instance \cite{sjostrom2018designing} have proposed a
conceptual architecture for the adoption of teaching-oriented chatbots in higher
education.  This conceptual architecture is based on a systematic literature
review of previous studies, examining the design of chatbots in higher
education, as well as making a content analysis of student emails and discussion
forum posts of four instances of a Java programming course.  The study outlined
several design considerations; among them, the authors emphasised the importance
of developing chatbots in platforms that students and educators are familiar
with and can easily access (i.e. Facebook Messenger) which was confirmed with
\cite{hobert2019you} and \cite{fernoagua2018intelligent} studies. In addition, \cite{sjostrom2018designing} argued that a
conceptualisation of learners' questions could aid designers in integrating the
appropriate types of questions that the chatbots should support for different
courses.  \textcolor{black}{Some other authors}, like \cite{coronado2018cognitive}, proposed agents that
store learning materials to be provided on demand to students, whereas the
authors in \cite{crockett2017predicting} reported on tutoring systems, which can
perform initial assessments of students' understanding and provide learning
material that would advance their understanding to the next level.

Regarding the factors for the adoption of chatbots in higher education, many
studies have focused on the evaluation of technology acceptance and usability
\cite{Roblyer2010,Pimmer2019}. However, higher education is a special domain
where, according to \cite{hobert2019you}, specific pedagogical factors
such as learning success and increased motivation are more important.  Therefore,
to develop \textcolor{black}{effective chatbots for higher education, all stakeholders'
(i.e. educators, students, institutions, etc) needs should be carefully
collected and taken into consideration
\cite{sjostrom2018designing,tsivitanidouusers}. These needs include, but are not limited to, student's learning success due to higher motivation, but these are a posteriori effects that cannot be assessed in advance; this is why we focus in this paper on what the teachers are looking for in a TAM  (Technology Acceptance Model) and associated technologies like chatbots; both authors focus on what the teachers need in terms of the latter; this paper will focus on a wider perspective trying to ascertain what they are looking for in terms of general messaging technology. The classical literature \cite{tam91} already proposes that any adoption of technology must be tuned in to the user needs and experience. In this paper we will try to find out what those are in order to propose a successful model of adoption of chatbot technology}.

Along this line of research, our previous work \cite{MoraChatbots2021}
aimed at analysing the expectations of students on this regard; recent papers
have also analysed how the use of chatbots affected the learning outcomes of
students in a Chinese class \cite{doi:10.1177/0735633120929622}. This paper
analysed conversational chatbots in an one-to-one setting, finding that learning
really benefited from it. This is, once again, a proof of the benefit of
chatbots in certain settings; however, there are some \textcolor{black}{prior experiences, as well as needs} that might
prevent the successful acceptance of the technology, and these are what we are
trying to find out in this paper, together with what kind of features would \textcolor{black}{improve
its acceptance}.

Besides, the present work is focused on the other key actor in this challenge,
i.e. educators. \textcolor{black}{How they accepted chatbots} has been studied very recently by Chocarro et
al. \cite{doi:10.1080/03055698.2020.1850426}, who, by analysing surveys, created
a {\em technology acceptance model} that proposed a series of features that
would make chatbots easier to accept; among them, formality of language as well
as easiness and usefulness. The survey targeted primary and secondary education
teachers, and was also more interested in the general use of chatbots in
education, not specifically in a classroom setting, seeking educational
outcomes. However, they establish a series of results that are obviously
interesting and relevant for this work.

We will present next the methodology we have followed to find out teachers'
opinions.


\section{Methodology}
\label{sec:meth}

\textcolor{black}{This study followed a quantitative approach towards addressing its research objective, by collecting rich data about the use of messaging applications and chatbots
by educators in both universities and colleges\footnote{In Spain, they are called simply high schools, and deliver
tertiary or associated degrees that do not have the consideration of university degrees}.
To achieve this goal two online surveys were designed and developed using Google Forms
in Spanish. They are comprised of six questions focusing on demographic data
(i.e., sector, gender, degrees, discipline, age, teaching experience)
followed by several multiple choice questions allowing the participants to choose multiple answers.}

In the {\bf first survey}, the multiple choice questions are
focused on the use of messaging apps in the teaching practice, the type of chatbot use cases that teachers consider
useful for their teaching and the impact of COVID on the teaching practice; please check them out to the full extent in the Appendix.\footnote{This {\bf first survey} was made in two separate Google forms, one for
university and the other for teachers in tertiary education institutions, outside the university system (vocational training, including 1- and 2-year degrees). Questions and
responses were the same, except for the type of tertiary degrees that
were considered.}

\textcolor{black}{This survey was piloted by the authors of the paper and their colleagues before using it in the 
study.  All the comments collected from this trial were incorporated in the survey.
We mainly used mailing lists, as well as Telegram groups,
to post the survey link to reach out to more educators. The university form was sent to Spanish
(mainly Andaluc\'{\i}a and Galicia) universities, and also  universities in Costa Rica and
Mexico. The tertiary (non-university/college) teachers were based mainly in
Andaluc\'{\i}a. Dissemination and answers took place in the first quarter of
2021, during the Covid-19 pandemic and while, at least in Spain, many universities had
mandatory virtual teaching.}

A total of 282 teachers responded to the survey: 193 teaching at the University
(68.4\%) and 89 at other tertiary education institutions (31.6\%). \textcolor{black}{From those, 179 indicated their gender as male (63.5\%), 98 as female (34.8\%) and 5 teachers preferred not to indicate their gender (1.8\%). In terms of age, the majority of the participants (n=111) in this survey was 45-55 years old (39.4\%), while 91 teachers were 35-45 (32.3\%), 46 were 25-35 (16.3\%) and 34 teachers were older than 55 (12.1\%). Last, 84 teachers had a teaching experience of 16-25 years (29.8\%), 75 teachers 6-15 years of experience (26.6\%), 69 teachers had 0-5 years of experience (24.5\%) and finally, 54 teachers had more than 25 years of teaching experience (19.1\%).}

Responses were stored automatically in a Google Drive spreadsheet. Eventually,
the results from the two forms we used for the first survey were collated in a
single spreadsheet. Survey questions are shown in Appendix I (Section \ref{app:questions}).

The {\bf second survey} was designed after initial results to the first survey arrived,
and pointed to necessities and experiences of teachers not being addressed by the first one; namely
their experience with messaging platforms and the way they were used to interact with students.
It was piloted in a group of university teachers attending to a formation course, and validated with them.
It was then extended to the rest of the responders, using the same media: Telegram groups, email and announcement in mailing lists. The questions asked in that survey are also shown in Appendix I
(Section \ref{app:questions}).

\textcolor{black}{A total of 205 teachers responded to the second survey: 187 graduate teachers (91.2\%) and 18 student teachers (8.8\%). From those, 124 were male (60.5\%), 65 female (31.7\%) and 16 teachers preferred not to indicate their gender (7.8\%). In terms of age, the majority of the participants (n=70) was again, as in the first survey, 45-55 years old (34.1\%), while 67 teachers were 35-45 years old (32.7\%), 42 were 25-35 (20.5\%) and 26 teachers were older than 55 (12.7\%). In terms of teaching experience, 59 teachers had a teaching experience of 16-25 years (28.8\%), 51 teachers had 6-15 years (24.9\%) and 0-5 years of experience (24.9\%), and 44 teachers more than 25 years of teaching experience (21.5\%).}

Next, we show the results and analyse them.


\section{Results and analysis}
\label{sec:res}

We collected responses from 282 teachers from Spain and Spanish Speaking
countries for \textit{survey1} and 205 for \textit{survey2}. The two forms were
open for approximately the same amount of time, around two months. The sample
includes mostly teachers from university, although about 32\% and 8\% are from
non-universitary tertiary education in the first and second survey respectively. Regarding
gender, 61\% of the teachers were male and 32\% female, while approximately 5\%
chose not to disclose it. Finally, the responses are more or less equally
distributed according to the teaching experience, showing 24\% of responses from
teachers with 5 or less years of experience, 26\% for 6-15 years of experience,
30\% for 16-25, and 20\% for teachers with more than 25 years of experience.

We will try to find the answers to the four research questions next, by
analysing the response to the different survey questions.


\subsection{RQ1 - Are teachers already using messaging apps in their classes?}


Group messaging apps have become a powerful tool for \textcolor{black}{communicating} with
students. Also, many of them have built-in chatbots to enhance the learning
process. After an initial analysis to the answers to the first survey, we
realized that there were some prior issues mainly related to the adoption of a
technology, \textcolor{black}{i.e. chatbots}, that generally piggybacks on another, \textcolor{black}{i.e. messaging
applications. Generally,} in a technology adoption model, perceived ease of use and
perceived usefulness are essential. But it is going to be very difficult for
teachers to find {\em chatbots} easy to use if they, previously, do not
consider, or simply use, the messaging tools to which they are attached.

This is why for answering RQ1, teachers were queried about whether they use
messaging apps in their classes, specifically, Telegram, WhatsApp, Slack (an
application used mainly in IT departments and software development), or any
other messaging app, and whether they use messaging apps provided by their
academic institution (see Table \ref{tab:use_messapps}).
Overall, the majority of the teachers responded that they do use messaging apps
in their classrooms, from which apps (or, more probably, messaging solutions) provided by the academic institution
(n=159, 56\%) and WhatsApp (n=124, 44.0\%) were the most common responses. Only
19 teachers (6.7\%) replied that they do not use any messaging app in their
class.
\begin{table}[h!tbp]
\begin{tabular}{lllll}
\hline
\multicolumn{1}{c}{Messaging App} & \multicolumn{2}{c}{Yes} & \multicolumn{2}{l}{No} \\ \cline{2-5} 
\multicolumn{1}{c}{} & \multicolumn{1}{c}{Frequency} & \multicolumn{1}{c}{\%} & Frequency & \multicolumn{1}{c}{\%} \\ \hline
Telegram & 62 & 22.0 & 220 & 78.0 \\
WhatsApp & 124 & 44.0 & 158 & 56.0 \\
Slack & 15 & 5.3 & 267 & 94.7 \\
Other & 60 & 21.3 & 222 & 78.7 \\
\begin{tabular}[c]{@{}l@{}}Provided by the Academic\\ Institution\end{tabular} & 159 & 56.4 & 123 & 43.6 \\
None & 19 & 6.7 & 263 & 93.3 \\ \hline
\end{tabular}
\vspace{0.8em}
\caption{Use of messaging apps to assist the learning process.}
\label{tab:use_messapps}
\end{table}

\begin{figure}[h!tbp]
\begin{center}
    \begin{minipage}[t]{0.35\textwidth}
    \centering
\begin{knitrout}
\definecolor{shadecolor}{rgb}{0.969, 0.969, 0.969}\color{fgcolor}
\includegraphics[width=\maxwidth]{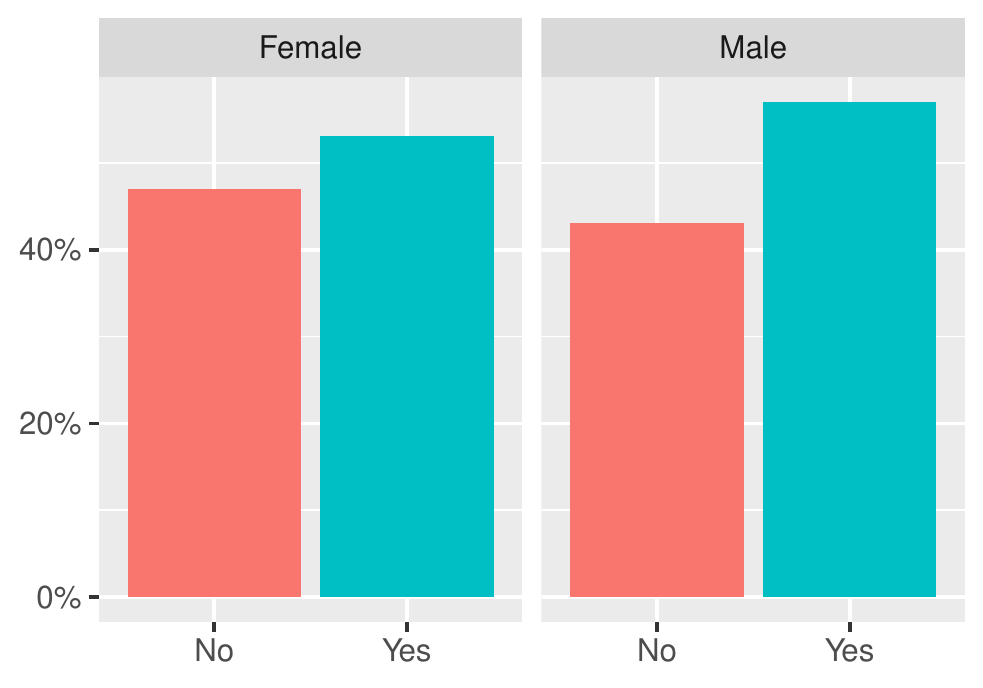} 

\end{knitrout}
    \end{minipage}
    \begin{minipage}[t]{0.35\textwidth}
    \centering
\begin{knitrout}
\definecolor{shadecolor}{rgb}{0.969, 0.969, 0.969}\color{fgcolor}
\includegraphics[width=\maxwidth]{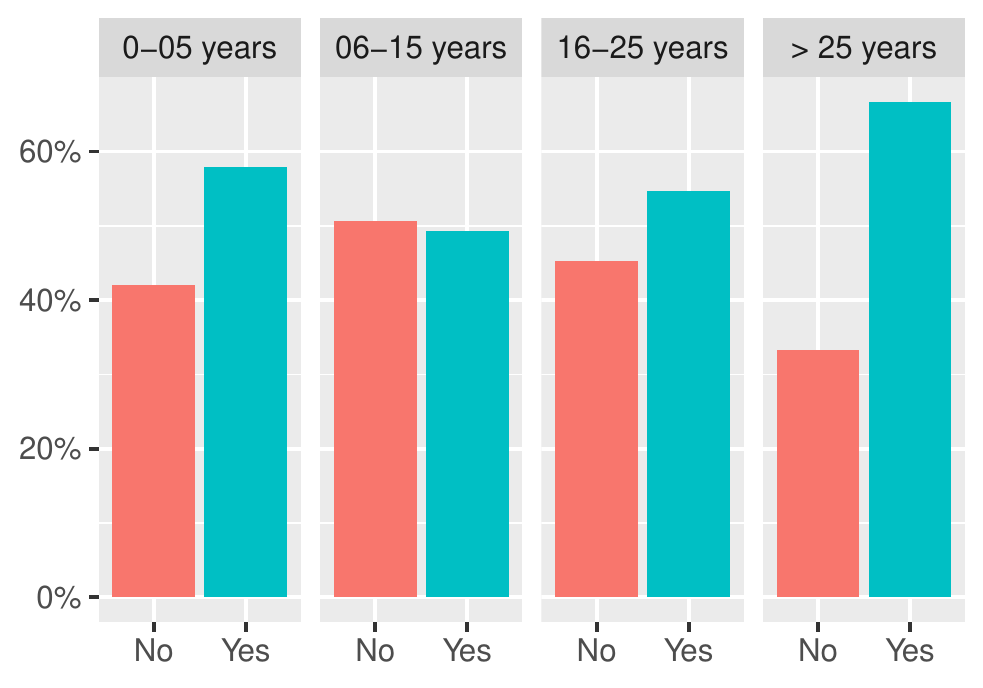} 

\end{knitrout}
    \end{minipage}
    \begin{minipage}[t]{0.3\textwidth}
        \centering
\begin{knitrout}
\definecolor{shadecolor}{rgb}{0.969, 0.969, 0.969}\color{fgcolor}
\includegraphics[width=\maxwidth]{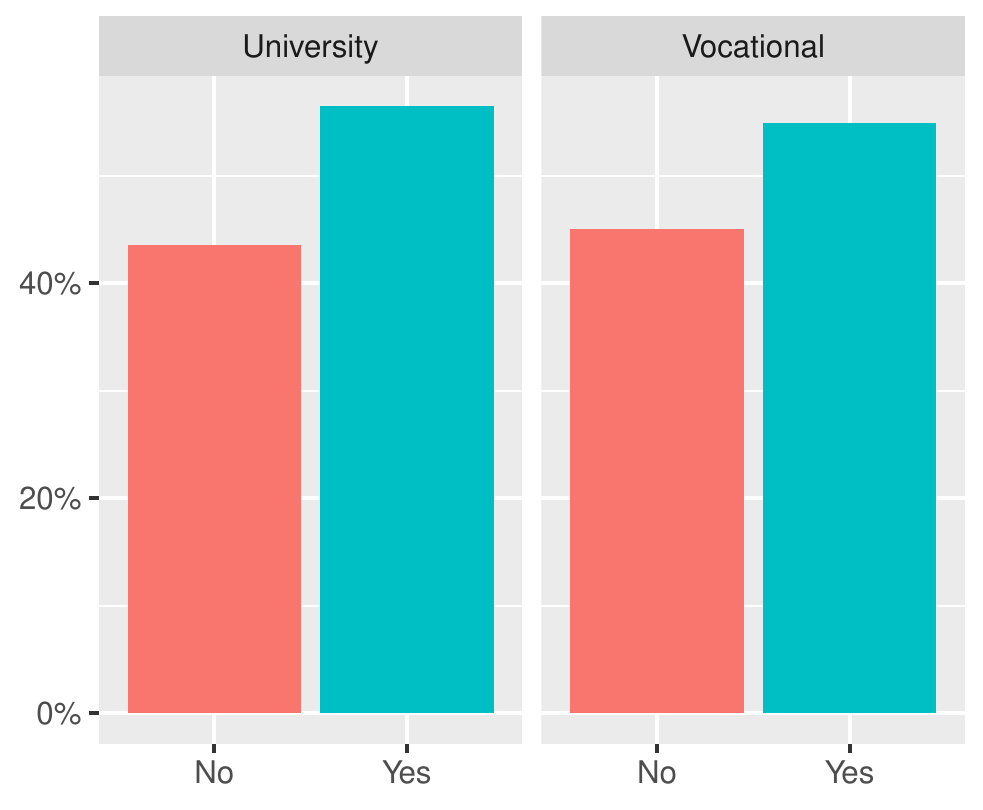} 

\end{knitrout}
    \end{minipage}
    \begin{minipage}[t]{0.6\textwidth}
        \centering
\begin{knitrout}
\definecolor{shadecolor}{rgb}{0.969, 0.969, 0.969}\color{fgcolor}
\includegraphics[width=\maxwidth]{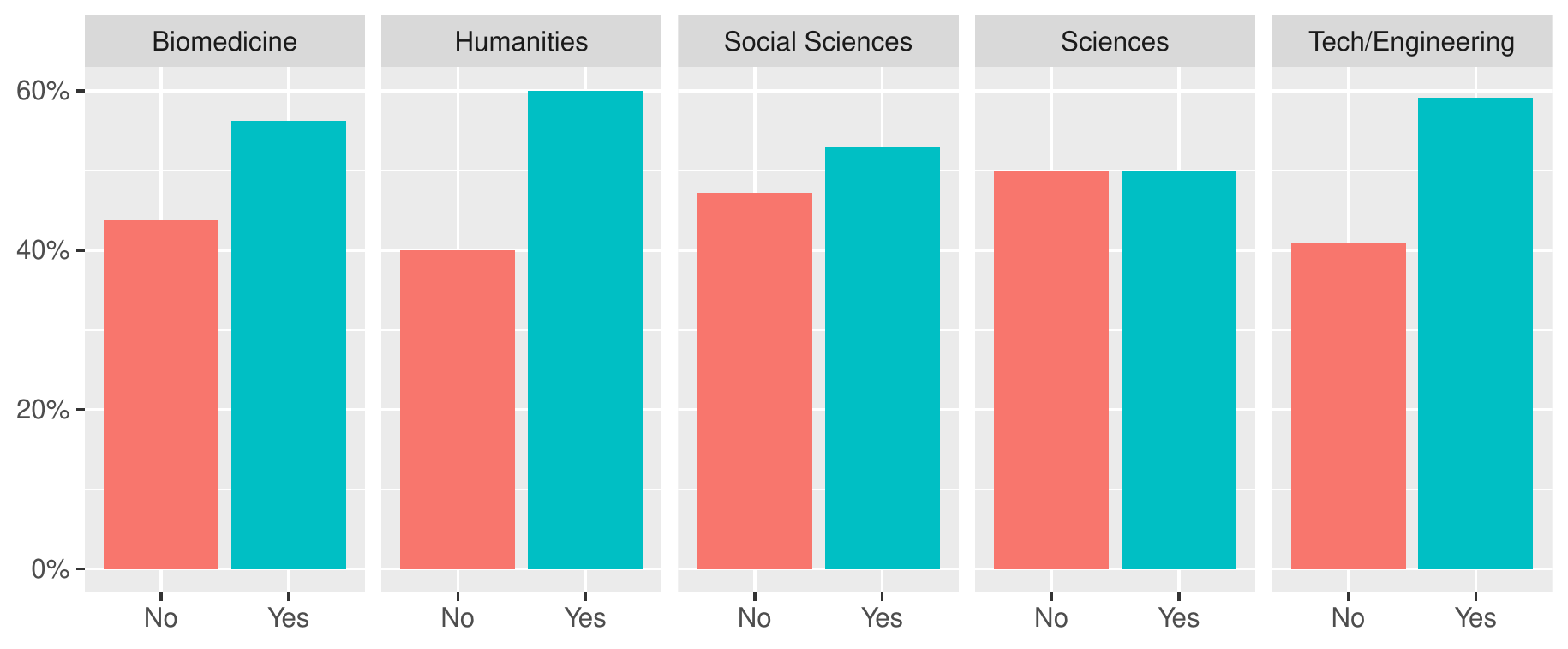} 

\end{knitrout}
    \end{minipage}
\end{center}
    \caption{Use of \textcolor{black}{messaging platforms either external or} provided by the teachers' academic institutions in class: distributions per gender, years of experience in education, university or vocational education, and discipline.}
    \label{figure:provided*chatbots_use}
\end{figure}

\begin{figure}[h!tbp]
\begin{center}
    \begin{minipage}[t]{0.35\textwidth}
    \centering
\begin{knitrout}
\definecolor{shadecolor}{rgb}{0.969, 0.969, 0.969}\color{fgcolor}
\includegraphics[width=\maxwidth]{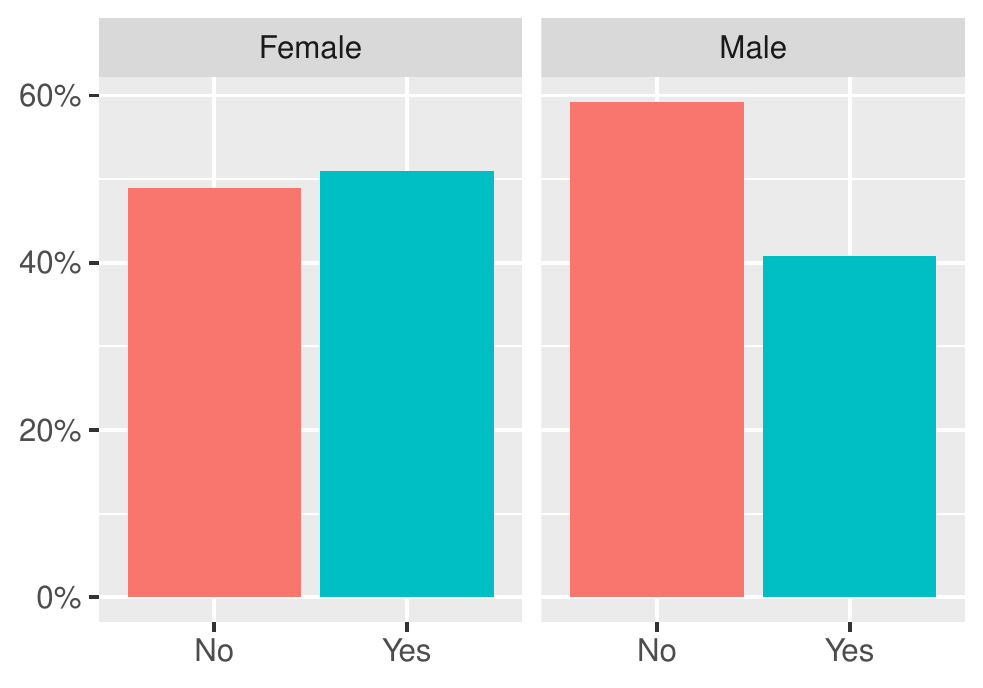} 

\end{knitrout}
    \end{minipage}
    \begin{minipage}[t]{0.35\textwidth}
    \centering
\begin{knitrout}
\definecolor{shadecolor}{rgb}{0.969, 0.969, 0.969}\color{fgcolor}
\includegraphics[width=\maxwidth]{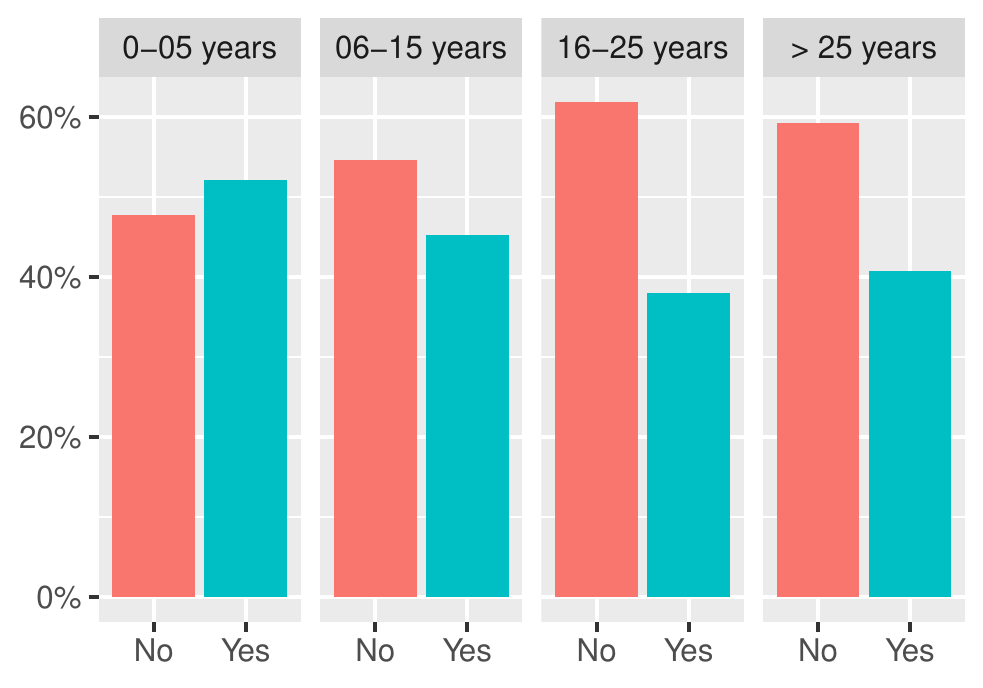} 

\end{knitrout}
    \end{minipage}
    \begin{minipage}[t]{0.3\textwidth}
    \centering
\begin{knitrout}
\definecolor{shadecolor}{rgb}{0.969, 0.969, 0.969}\color{fgcolor}
\includegraphics[width=\maxwidth]{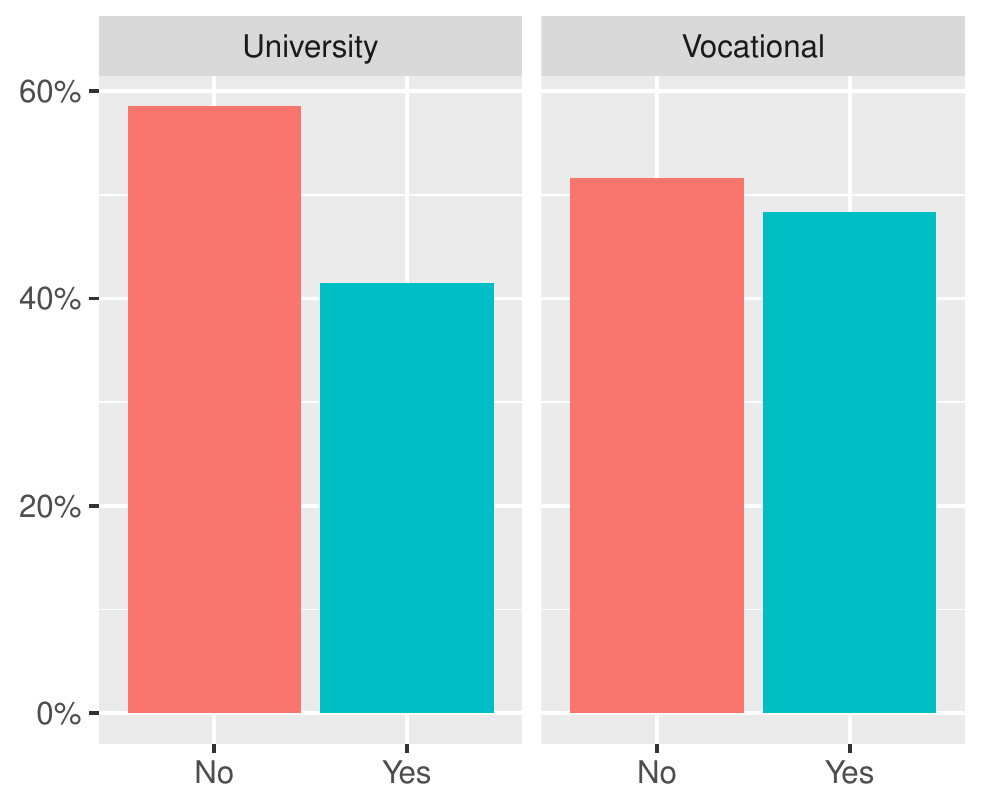} 

\end{knitrout}
    \end{minipage}
    \begin{minipage}[t]{0.6\textwidth}
    \centering
\begin{knitrout}
\definecolor{shadecolor}{rgb}{0.969, 0.969, 0.969}\color{fgcolor}
\includegraphics[width=\maxwidth]{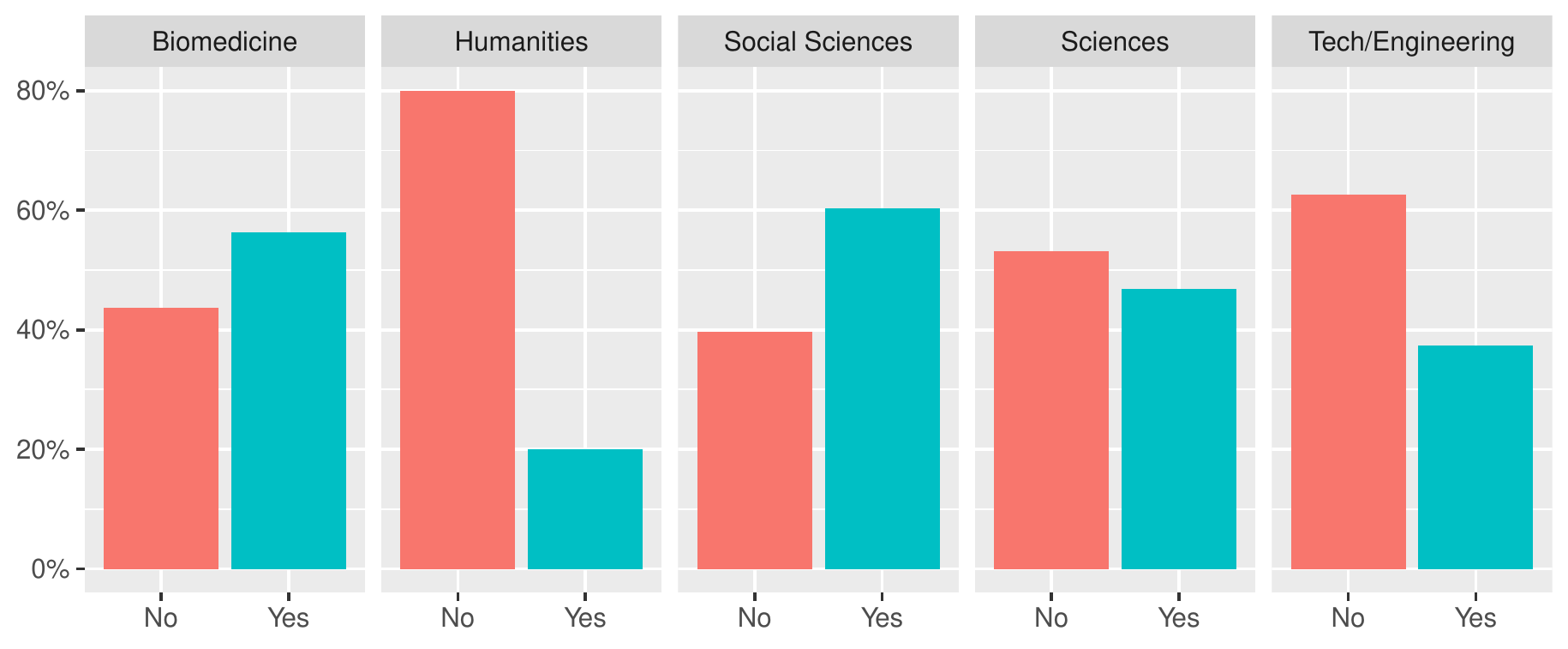} 

\end{knitrout}
    \end{minipage}
\end{center}
    \caption{Use of Whatsapp in class: distributions per gender, years of experience in education, university or vocational education, and discipline.}
    \label{figure:whatsapp*chatbots_use}
\end{figure}

\begin{figure}[h!tbp]
\begin{center}
    \begin{minipage}[t]{0.35\textwidth}
    \centering
\begin{knitrout}
\definecolor{shadecolor}{rgb}{0.969, 0.969, 0.969}\color{fgcolor}
\includegraphics[width=\maxwidth]{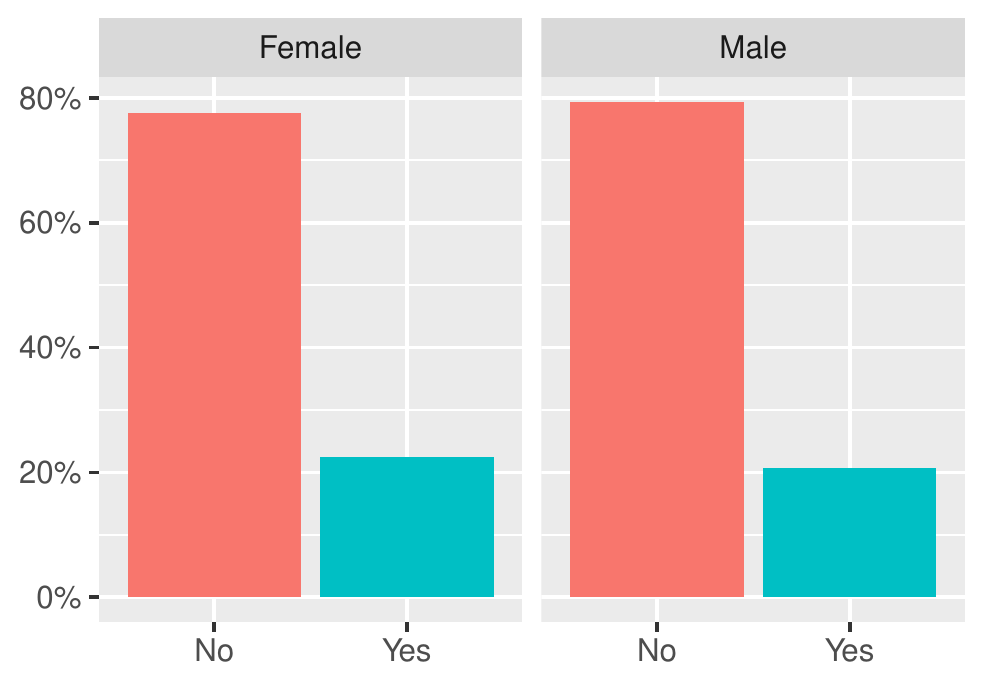} 

\end{knitrout}
    \end{minipage}
    \begin{minipage}[t]{0.35\textwidth}
    \centering
\begin{knitrout}
\definecolor{shadecolor}{rgb}{0.969, 0.969, 0.969}\color{fgcolor}
\includegraphics[width=\maxwidth]{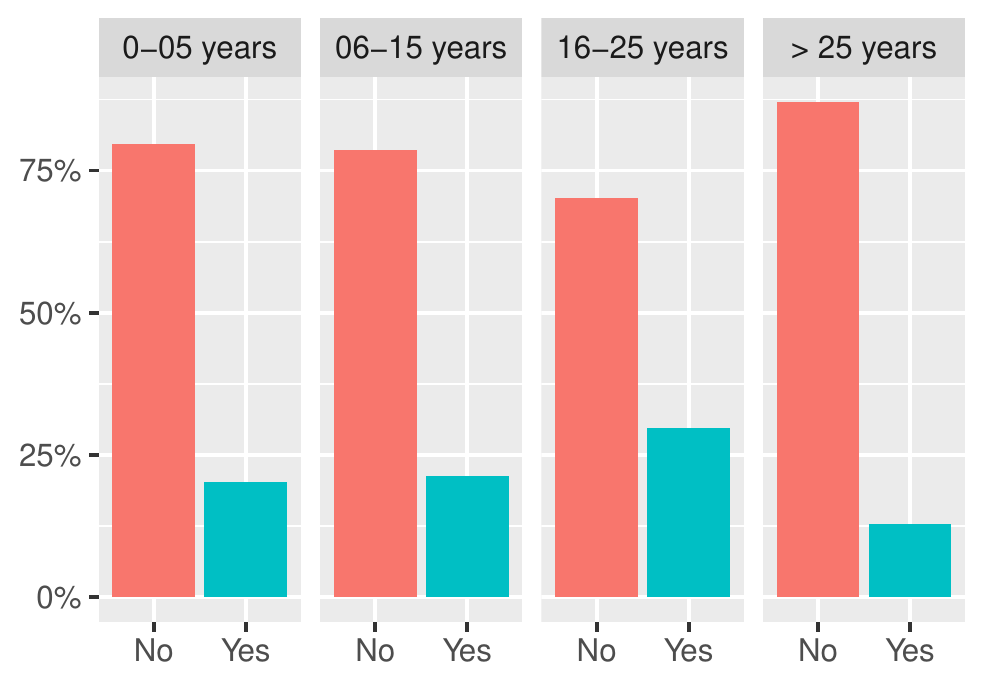} 

\end{knitrout}
    \end{minipage}
    \begin{minipage}[t]{0.3\textwidth}
    \centering
\begin{knitrout}
\definecolor{shadecolor}{rgb}{0.969, 0.969, 0.969}\color{fgcolor}
\includegraphics[width=\maxwidth]{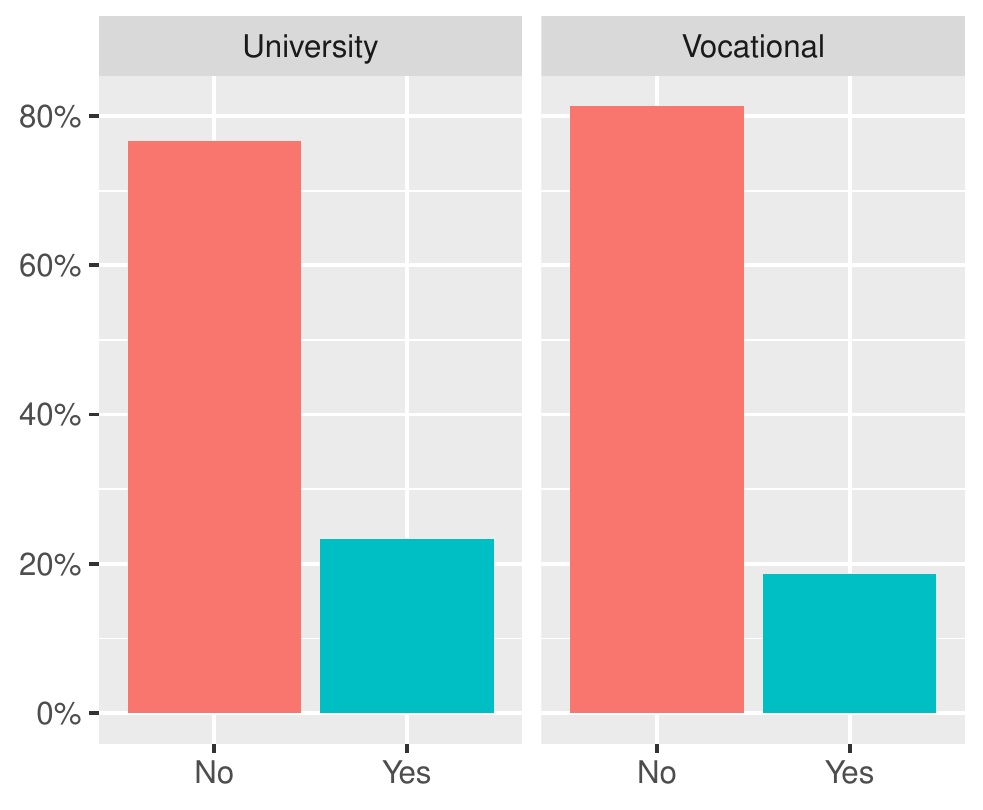} 

\end{knitrout}
    \end{minipage}
    \begin{minipage}[t]{0.6\textwidth}
    \centering
\begin{knitrout}
\definecolor{shadecolor}{rgb}{0.969, 0.969, 0.969}\color{fgcolor}
\includegraphics[width=\maxwidth]{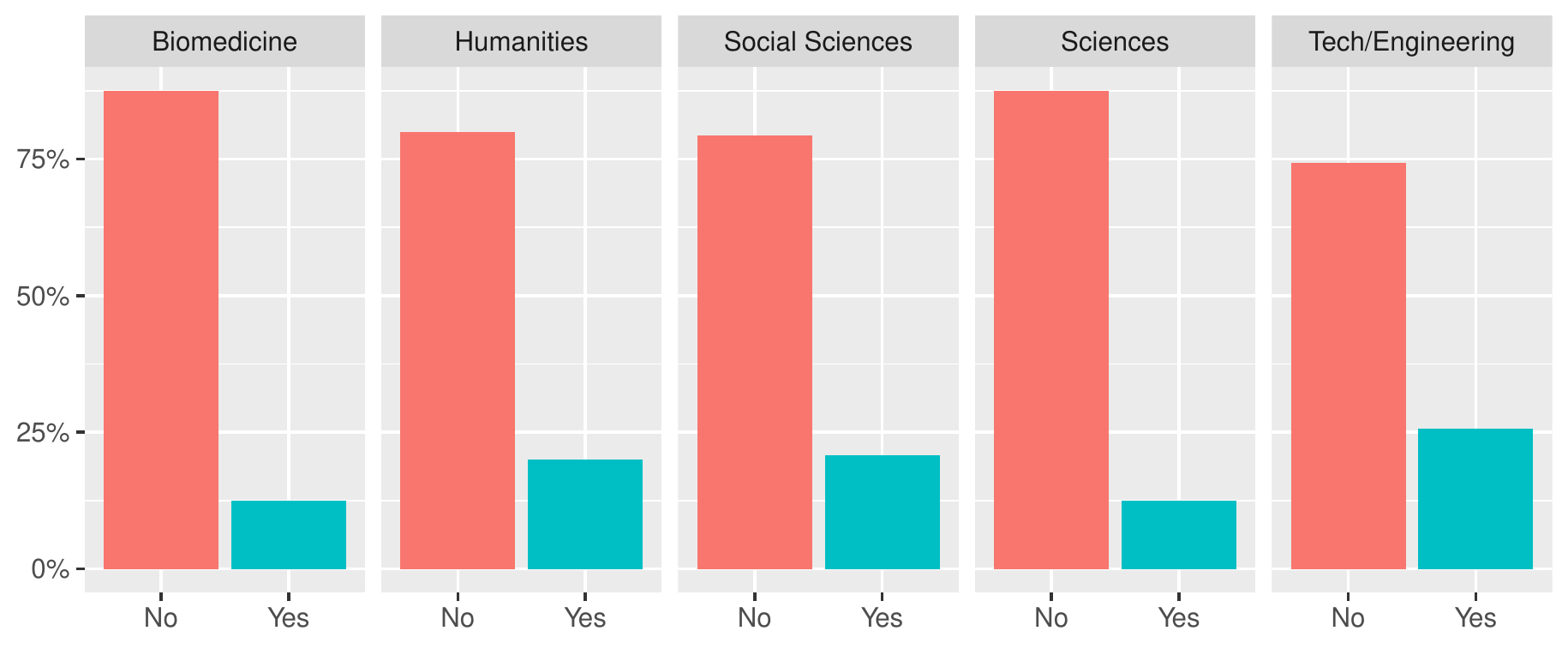} 

\end{knitrout}
    \end{minipage}
\end{center}
    \caption{Use of Telegram in class: distributions per gender, years of experience in education, university or vocational education, and discipline.}
    \label{figure:telegram*chatbots_use}
\end{figure}

Results in Figs. \ref{figure:provided*chatbots_use},
\ref{figure:whatsapp*chatbots_use}, \ref{figure:telegram*chatbots_use} do not
show significant differences in the use of instant messaging apps between
teachers from University and vocational education. In general, most teachers
prefer messaging apps provided by their own institutions (we also show the
results for the two most popular messaging app platforms: WhatsApp and
Telegram). With respect to specific disciplines, Engineering and Technology
teachers are more active in their use, \textcolor{black}{but the number of teachers from Humanities 
who answered they used these apps in their classes is also remarkable}
(around 60\% use the apps provided by their institutions).

Although no significant differences were found regarding gender, female teachers
answered they use instant messaging apps more than male teachers (about 10\%
more). Also, teachers in vocational Education use WhatsApp more than university
teachers. Regarding the distribution of the use of messaging apps per age, there
are no significant differences for WhatsApp and apps provided by their own
academic institutions. 
However, younger teachers also use Telegram with more than 25\% responding they do, a percentage
that falls to about 10\% for teachers that are 55 or older. 
One interesting result is that about 65\% of teachers with more than 25 years of
experience use the platforms provided by their institutions while the percentage
goes down to less than 50\% for teachers with 6-15 years of experience.

\begin{figure}[h!tbp]
\begin{center}
    \begin{minipage}[t]{0.80\textwidth}
        \centering
\begin{knitrout}
\definecolor{shadecolor}{rgb}{0.969, 0.969, 0.969}\color{fgcolor}
\includegraphics[width=\maxwidth]{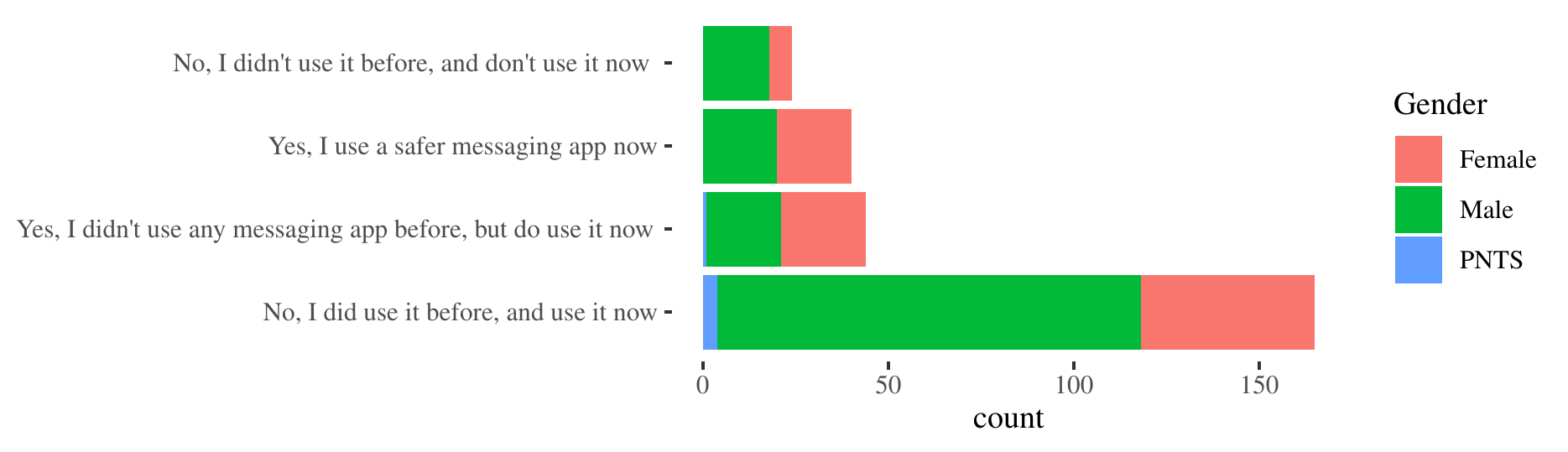} 

\end{knitrout}
        \vspace{-0.5cm}
        \caption{Total count of responses for the use of messaging apps after the COVID-19 pandemic grouped by gender (PNTS stands for Prefer Not To Say).}
        \label{figure:gender*chatbots_use_postcovid}
    \end{minipage}
    \begin{minipage}[t]{0.80\textwidth}
        \centering
\begin{knitrout}
\definecolor{shadecolor}{rgb}{0.969, 0.969, 0.969}\color{fgcolor}
\includegraphics[width=\maxwidth]{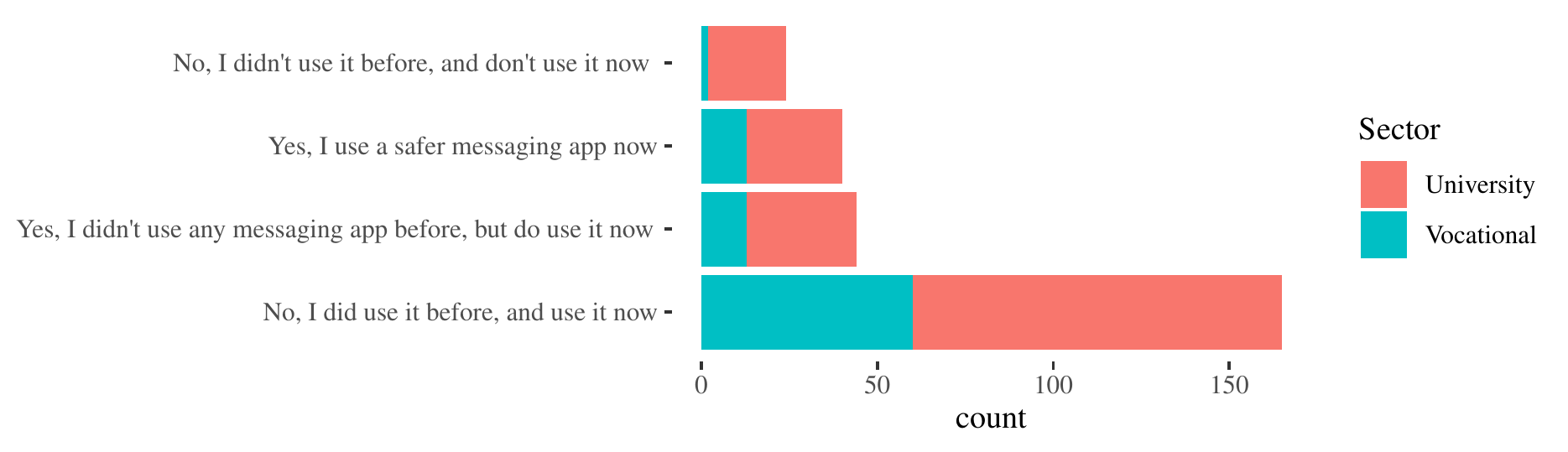} 

\end{knitrout}
        \vspace{-0.5cm}
        \caption{Total count of responses for the use of messaging apps after the COVID-19 pandemic grouped by sector.}
        \label{figure:sector*chatbots_use_postcovid}
    \end{minipage}
    \begin{minipage}[t]{0.80\textwidth}
        \centering
\begin{knitrout}
\definecolor{shadecolor}{rgb}{0.969, 0.969, 0.969}\color{fgcolor}
\includegraphics[width=\maxwidth]{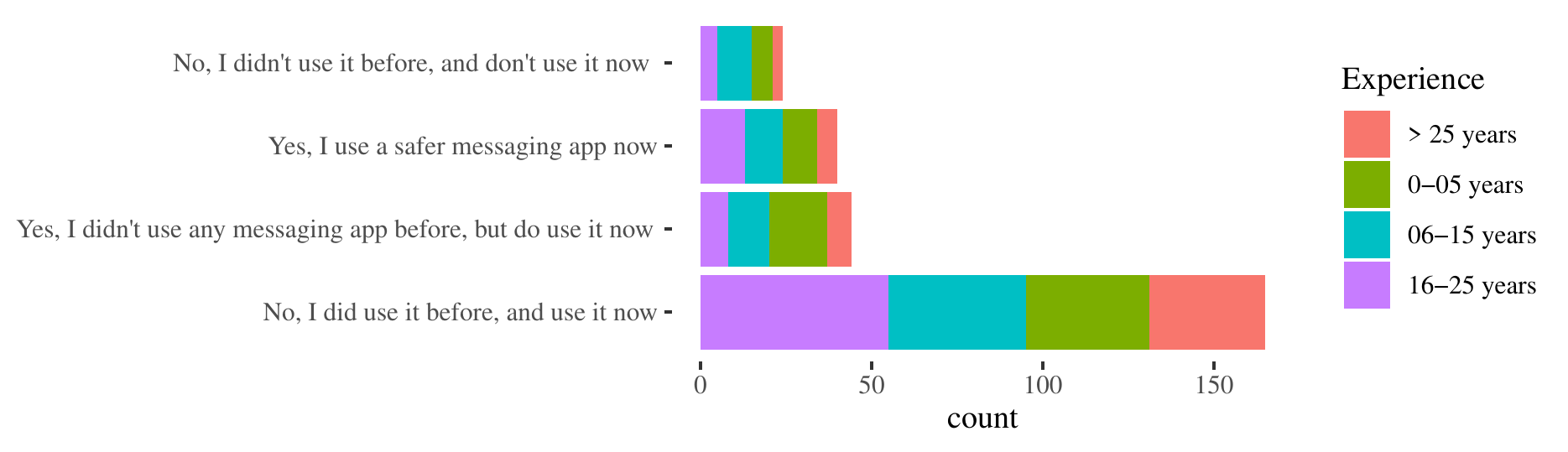} 

\end{knitrout}
        \vspace{-0.5cm}
        \caption{Total count of responses for the use of messaging apps after the COVID-19 pandemic grouped by years of experience in teaching.}
        \label{figure:years*chatbots_use_postcovid}
    \end{minipage}
        \begin{minipage}[t]{0.80\textwidth}
        \centering
\begin{knitrout}
\definecolor{shadecolor}{rgb}{0.969, 0.969, 0.969}\color{fgcolor}
\includegraphics[width=\maxwidth]{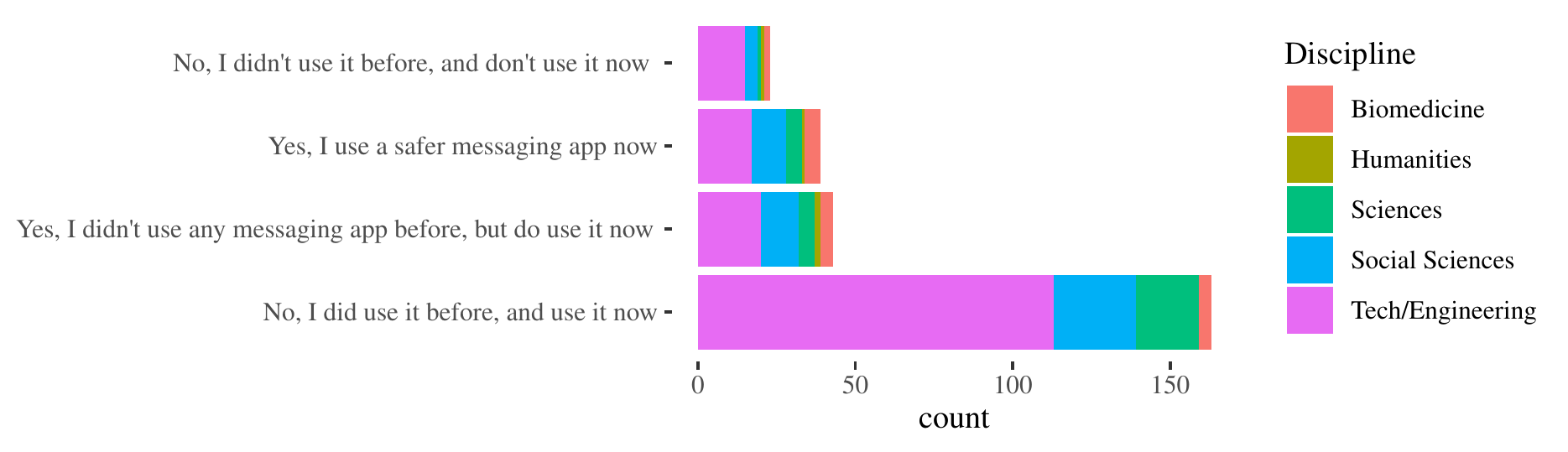} 

\end{knitrout}
        \vspace{-0.5cm}
        \caption{Total count of responses for the use of messaging apps after the COVID-19 pandemic grouped by discipline.}
        \label{figure:disc*chatbots_use_postcovid}
    \end{minipage}
\end{center}
\end{figure}

Some of the questions in the survey were focused on the impact of the COVID19
pandemic between the 2020 and 2021 academic years in the teachers' attitudes towards the use of
instant messaging apps in their class. Our main intention in this case was to
assess whether a crisis will bring about some kind of change in the use of
tools. The 282 answers are summarized in
Figs. \ref{figure:gender*chatbots_use_postcovid},
\ref{figure:sector*chatbots_use_postcovid},
\ref{figure:years*chatbots_use_postcovid},
\ref{figure:disc*chatbots_use_postcovid}, showing that about 77\% of teachers
already used these tools before the pandemic and kept using them during the
pandemic lockdowns that forced students and educators to use remote education
schemes. Moreover, approximately 15\% of them switched their messaging app for
one that offered a safer interaction with their students. According to the
responses, an additional 16\% started using messaging apps during the pandemic
for the first time in their classes. 

A chi-square test of independence was performed to examine the relation between instructors’ discipline, sector, and gender and any potential changes that occurred to the use of messaging apps due to the covid-19 pandemic. The relation between the latter variable and the instructors’ sector was significant, $X^{2}_{(4, N = 282)} = 9.598$, $p =0.048$. The frequencies cross tabulated are given in Table \ref{tab:sector*covid}. In effect, this finding indicates that how teachers responded to the use of messaging apps during the pandemic and in particular, whether they changed their habits in the use of apps for teaching purposes, was related to their sector (university vs vocational).
\begin{table}[h!tbp]
\begin{tabular}{llccc}
\hline
 & \multicolumn{1}{c}{} & \multicolumn{2}{c}{Sector} & Total \\ \cline{3-5} 
 & \multicolumn{1}{c}{} & \begin{tabular}[c]{@{}c@{}}University \\ (f)\end{tabular} & \begin{tabular}[c]{@{}c@{}}Vocational\\ (f)\end{tabular} &  \\ \cline{2-5} 
\multirow{4}{*}{\begin{tabular}[c]{@{}l@{}}Post-covid\\ changes\end{tabular}} & \begin{tabular}[c]{@{}l@{}}Yes, I use a safer messaging \\ app now\end{tabular} & 27 & 13 & 40 \\
 & \begin{tabular}[c]{@{}l@{}}Yes, I didn’t use any \\ messaging app before, but \\ do use it now\end{tabular} & 31 & 13 & 44 \\
 & \begin{tabular}[c]{@{}l@{}}No, I did use it before and \\ use it now\end{tabular} & 105 & 60 & 165 \\
 & \begin{tabular}[c]{@{}l@{}}No, I didn’t use it before and \\ don’t use it now\end{tabular} & 22 & 2 & 24 \\
Total &  & 193 & 89 & 282 \\ \hline
\end{tabular}
\vspace{0.8em}
\caption{Sector * post-covid changes cross-tabulation}
\label{tab:sector*covid}
\end{table}

\begin{figure}[h!tbp]
\begin{center}
    \begin{minipage}[c]{0.45\textwidth}
        \centering
\begin{knitrout}
\definecolor{shadecolor}{rgb}{0.969, 0.969, 0.969}\color{fgcolor}
\includegraphics[width=\maxwidth]{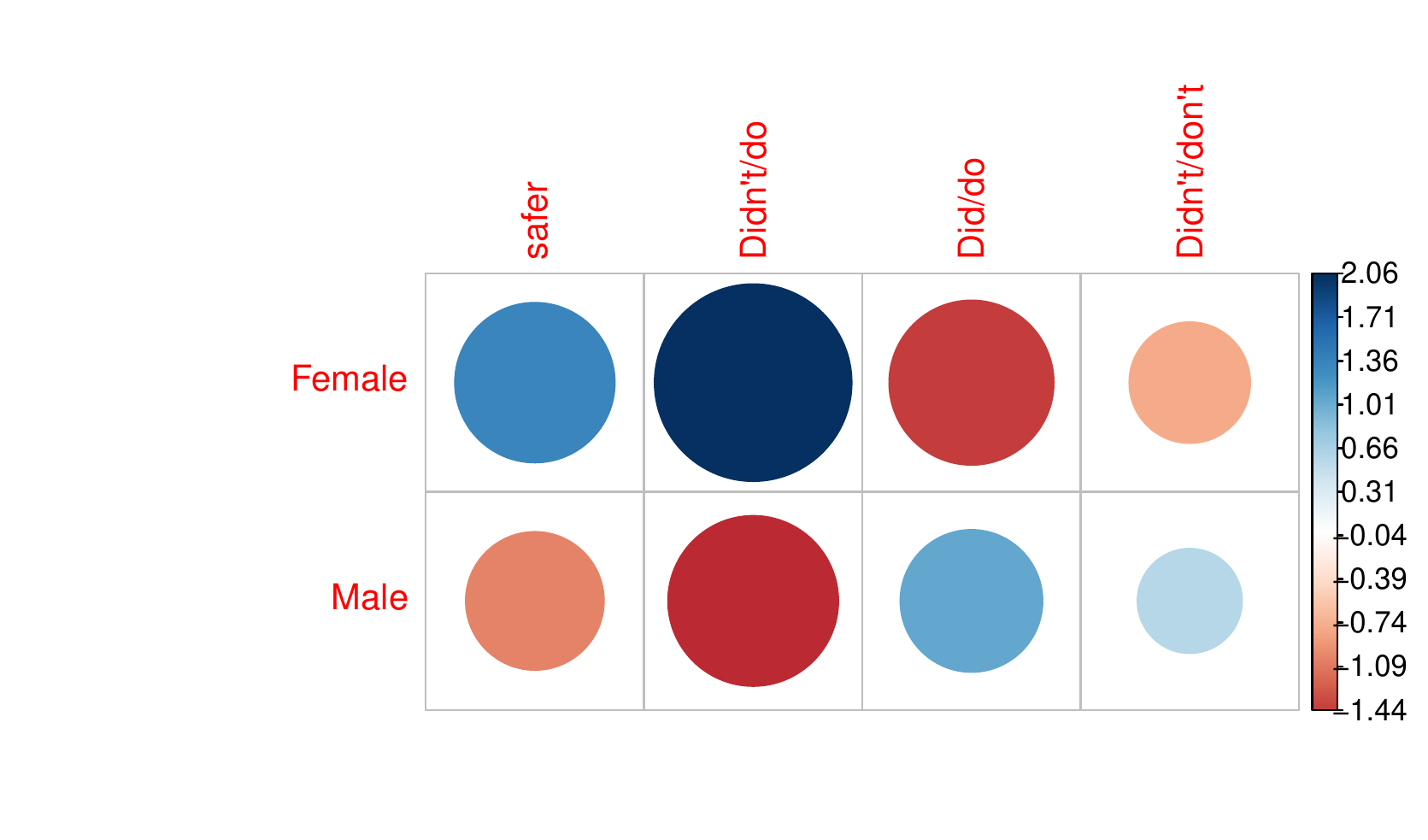} 

\end{knitrout}
    \end{minipage}
    \begin{minipage}[c]{0.50\textwidth}
        \centering
    \resizebox{\textwidth}{!}{%
\begin{tabular}{llllll}
\hline
 &  & \multicolumn{4}{c}{Gender} \\ \cline{3-6} 
 &  & \multicolumn{2}{c}{Female} & \multicolumn{2}{c}{Male} \\ \cline{3-6} 
 &  & \multicolumn{1}{c}{\begin{tabular}[c]{@{}c@{}}Stand. \\ Residuals\end{tabular}} & \multicolumn{1}{c}{\begin{tabular}[c]{@{}c@{}}Contri-\\ bution\end{tabular}} & \multicolumn{1}{c}{\begin{tabular}[c]{@{}c@{}}Stand. \\ Residual\end{tabular}} & \multicolumn{1}{c}{\begin{tabular}[c]{@{}c@{}}Contri-\\ bution\end{tabular}} \\ \cline{2-6} 
\begin{tabular}[c]{@{}l@{}}Post-\\ covid\\ changes\end{tabular} & \begin{tabular}[c]{@{}l@{}}Yes, I use a \\ safer messaging \\ app now\end{tabular} & 1.832 & 13.49\% & -1.832 & 7.50\% \\
 & \begin{tabular}[c]{@{}l@{}}Yes, I didn’t use \\ any messaging \\ app before, but\\ do use it now\end{tabular} & \textbf{2.806} & \textbf{31.21\%} & \textbf{-2.806} & \textbf{17.36\%} \\
 & \begin{tabular}[c]{@{}l@{}}No, I did use it \\ before and\\ use it now\end{tabular} & \textbf{2.850} & 15.15\% & \textbf{-2.850} & 8.43\% \\
 & \begin{tabular}[c]{@{}l@{}}No, I didn’t use \\ it before and\\ don’t use it now\end{tabular} & -1.011 & 4.40\% & 1.011 & 2.45\% \\ \hline
\end{tabular}
    }
    \end{minipage}
\end{center}
    \caption{Analysis for gender and post-covid changes: left) Pearson's residuals for the Chi-square test; right) Gender*post-covid changes crosstabulation, including standardized residuals \cite{agresti2013categorical} (also adjusted standardized residuals) and contribution percentage to the total Chi-square test of each cell.}
\label{figure:gender*postcovid_chisq}
\end{figure}

\begin{figure}[h!tbp]
\begin{center}
    \begin{minipage}[c]{0.45\textwidth}
        \centering
\begin{knitrout}
\definecolor{shadecolor}{rgb}{0.969, 0.969, 0.969}\color{fgcolor}
\includegraphics[width=\maxwidth]{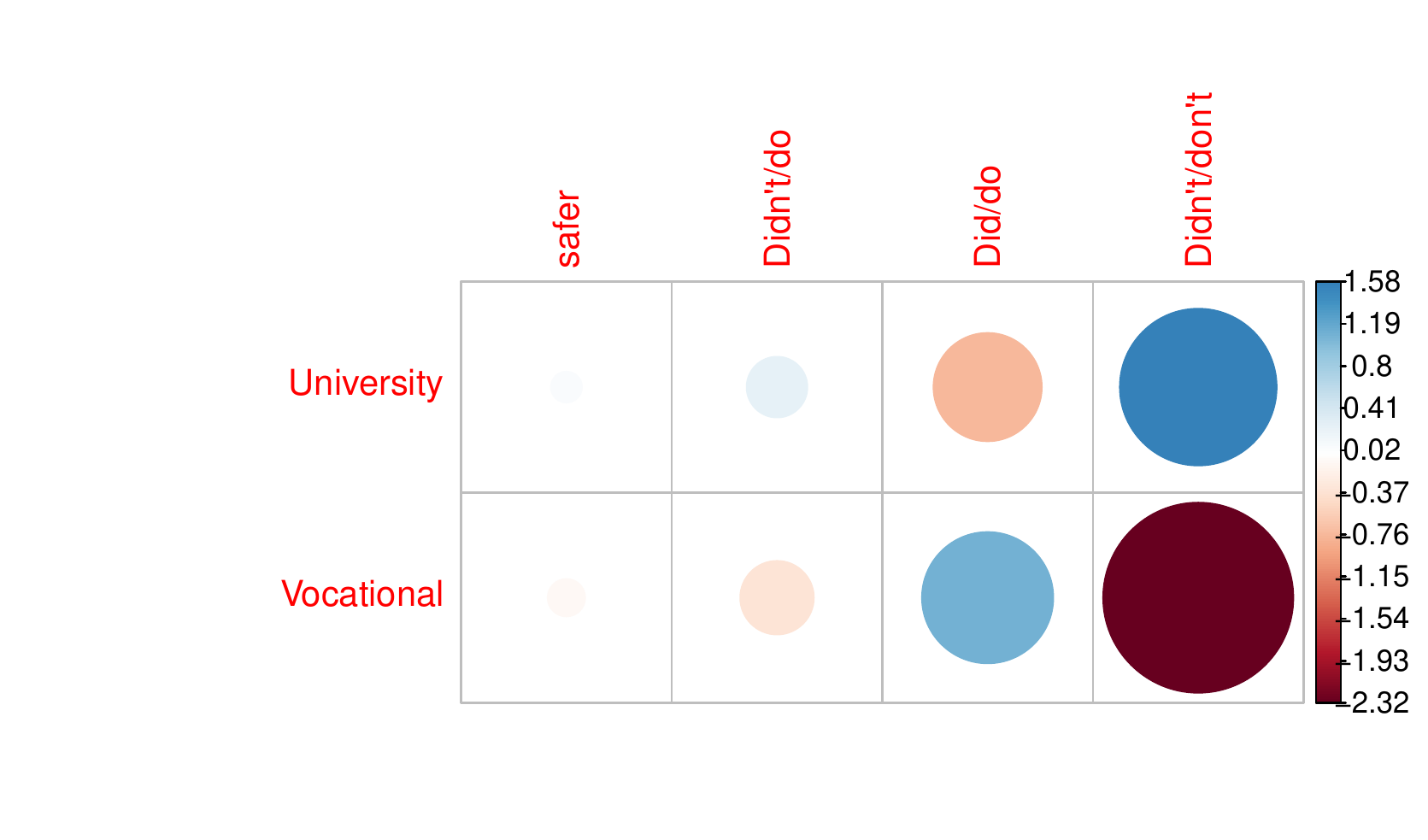} 

\end{knitrout}
    \end{minipage}
    \begin{minipage}[c]{0.50\textwidth}
        \centering
    \resizebox{\textwidth}{!}{%
    \begin{tabular}{llllll}
          \hline
           &  & \multicolumn{4}{c}{Sector} \\ \cline{3-6} 
           &  & \multicolumn{2}{c}{University} & \multicolumn{2}{c}{Vocational} \\ \cline{3-6} 
           &  & \multicolumn{1}{c}{\begin{tabular}[c]{@{}c@{}}Stand. \\ Residuals\end{tabular}} & \multicolumn{1}{c}{\begin{tabular}[c]{@{}c@{}}Contri-\\ bution\end{tabular}} & \multicolumn{1}{c}{\begin{tabular}[c]{@{}c@{}}Stand. \\ Residual\end{tabular}} & \multicolumn{1}{c}{\begin{tabular}[c]{@{}c@{}}Contri-\\ bution\end{tabular}} \\ \cline{2-6} 
          \begin{tabular}[c]{@{}l@{}}Post-\\ covid\\ changes\end{tabular} & \begin{tabular}[c]{@{}l@{}}Yes, I use a \\ safer messaging \\ app now\end{tabular} & 0.115 & 0.04\% & -0.115 & 0.08\% \\
           & \begin{tabular}[c]{@{}l@{}}Yes, I didn’t use \\ any messaging \\ app before, but\\ do use it now\end{tabular} & 0.454 & 0.56\% & -0.454 & 1.21\% \\
           & \begin{tabular}[c]{@{}l@{}}No, I did use it \\ before and\\ use it now\end{tabular} & \textbf{-2.122} & 5.71\% & \textbf{2.122} & 12.38\% \\
           & \begin{tabular}[c]{@{}l@{}}No, I didn’t use \\ it before and\\ don’t use it now\end{tabular} & \textbf{2.939} & \textbf{25.26\%} & \textbf{-2.939} & \textbf{54.78\%} \\ \hline
          \end{tabular}%
    }
    \end{minipage}
\end{center}
    \caption{Analysis for sector and post-covid changes: left) Pearson's residuals for the Chi-square test; right) Sector*post-covid changes crosstabulation, including standardized residuals \cite{agresti2013categorical} (also adjusted standardized  residuals) and contribution percentage to the total Chi-square test of each cell.}
\label{figure:sector*postcovid_chisq}
\end{figure}

\begin{figure}[h!tbp]
\begin{center}
    \begin{minipage}[c]{0.45\textwidth}
        \centering
\begin{knitrout}
\definecolor{shadecolor}{rgb}{0.969, 0.969, 0.969}\color{fgcolor}
\includegraphics[width=\maxwidth]{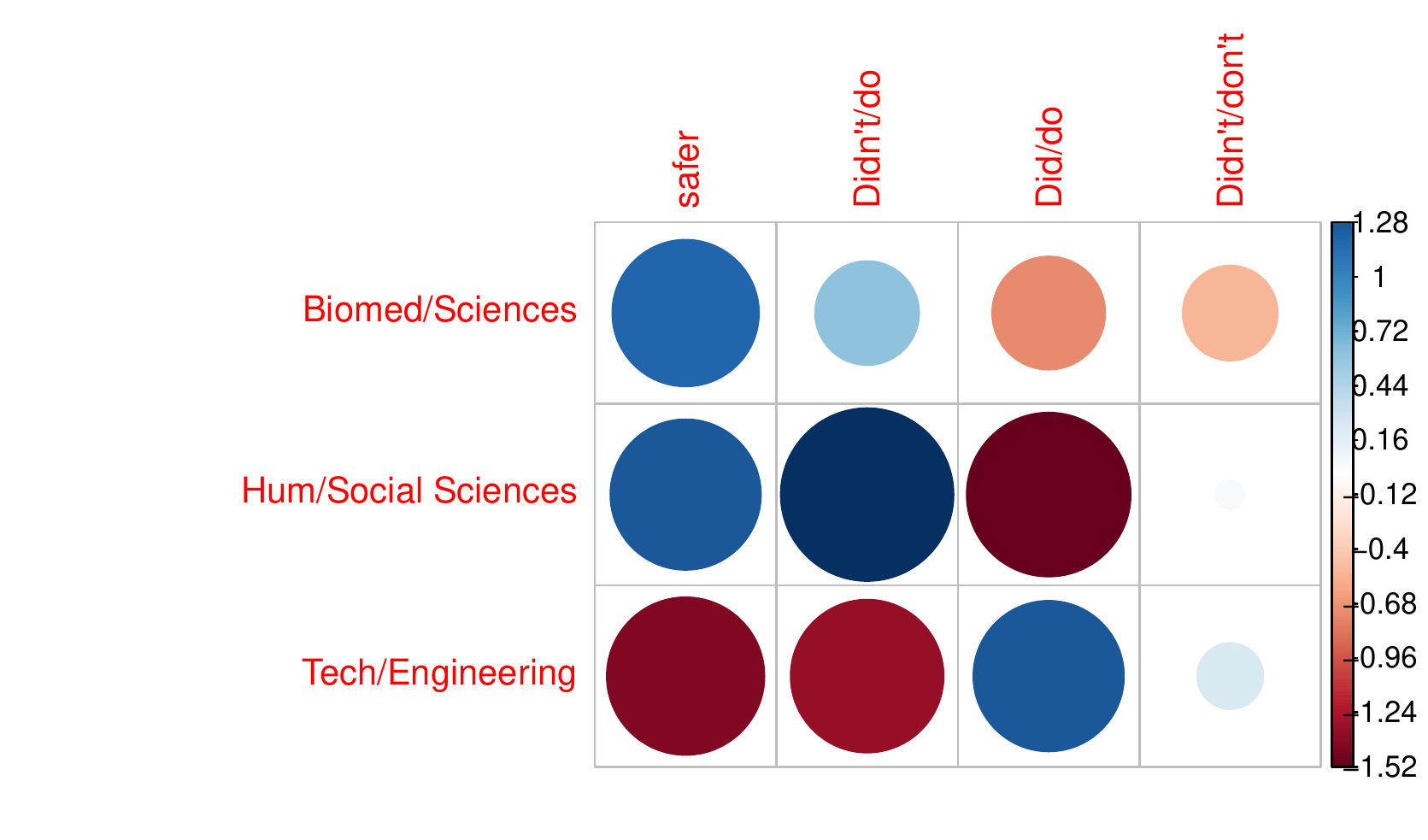} 

\end{knitrout}
    \end{minipage}
    \begin{minipage}[c]{0.50\textwidth}
        \centering
    \resizebox{\textwidth}{!}{%
\begin{tabular}{llllllll}
\hline
 &  & \multicolumn{6}{c}{Discipline} \\ \cline{3-8} 
 &  & \multicolumn{2}{c}{Biomed/Sciences} & \multicolumn{2}{c}{Hum/Social} & \multicolumn{2}{l}{Tech/Engineering} \\ \cline{3-8} 
 &  & \multicolumn{1}{c}{\begin{tabular}[c]{@{}c@{}}Stand. \\ Residuals\end{tabular}} & \multicolumn{1}{c}{\begin{tabular}[c]{@{}c@{}}Contri-\\ bution\end{tabular}} & \multicolumn{1}{c}{\begin{tabular}[c]{@{}c@{}}Stand. \\ Residual\end{tabular}} & \multicolumn{1}{c}{\begin{tabular}[c]{@{}c@{}}Contri-\\ bution\end{tabular}} & \begin{tabular}[c]{@{}l@{}}Stand. \\ Residuals\end{tabular} & \begin{tabular}[c]{@{}l@{}}Contri-\\ bution\end{tabular} \\ \cline{2-8} 
\begin{tabular}[c]{@{}l@{}}Post-\\ covid\\ changes\end{tabular} & \begin{tabular}[c]{@{}l@{}}Yes, I use a \\ safer mes-\\ saging app \\ now\end{tabular} & 1.452 & \textbf{9.94\%} & 1.566 & 11.04\% & \textbf{-2.448} & \textbf{13.29\%} \\
 & \begin{tabular}[c]{@{}l@{}}Yes, I didn’t \\ use any mes-\\ saging app \\ before, but\\ do use it now\end{tabular} & 0.733 & 2.50\% & \textbf{2.079} & \textbf{19.19\%} & -2.318 & 11.76\% \\
 & \begin{tabular}[c]{@{}l@{}}No, I did use \\ it before and\\ use it now\end{tabular} & -1.265 & 3.50\% & \textbf{-2.729} & 15.55\% & \textbf{3.278} & 11.06\% \\
 & \begin{tabular}[c]{@{}l@{}}No, I didn’t \\ use it before \\ and don’t use \\ it now\end{tabular} & -0.588 & 1.75\% & 0.055 & 0.01\% & 0.412 & 0.40\% \\ \hline
\end{tabular}
    }
    \end{minipage}
\end{center}
    \caption{Analysis for discipline and post-covid changes: left) Pearson's residuals for the Chi-square test; right) Discipline*post-covid changes crosstabulation, including standardized residuals \cite{agresti2013categorical} (also adjusted standardized residuals) and contribution percentage to the total Chi-square test of each cell.}
\label{figure:discipline*postcovid_chisq}
\end{figure}

The majority of the teachers (n=165) mentioned that no changes in their habits occurred due to the emergency remote teaching, as the use of messaging apps was part of their teaching practices and remains the same, from which 105 teachers come from the university, and 60 teachers from the non-universitary tertiary sector. The relationships between changes in the use of messaging apps, due to the covid-19 pandemic, and the instructors’ discipline, $X^{2}_{(24, N = 282)} = 44.856$, $p =0.006$, as well as the gender, $X^{2}_{(8, N = 282)} = 16.249$, p= 0.039 were also significant. This finding indicates that how teachers responded to the use of messaging apps during the pandemic was also related to their gender and discipline. In fact, from the majority of the teachers who did not change their habits in this respect (n=165), most of them are males (n=114), and come from the technology (n=60) and engineering (n=53) disciplines (see Tables \ref{tab:gender*covid}, \ref{tab:discipline*covid}).
\begin{table}[h!tbp]
\begin{tabular}{llcccc}
\hline
 & \multicolumn{1}{c}{} & \multicolumn{3}{c}{Gender} & Total \\ \cline{3-6} 
 & \multicolumn{1}{c}{} & \begin{tabular}[c]{@{}c@{}}Male \\ (f)\end{tabular} & \begin{tabular}[c]{@{}c@{}}Female\\ (f)\end{tabular} &  \begin{tabular}[c]{@{}c@{}}PNTS\\ (f)\end{tabular} & \\ \cline{2-6} 
\multirow{4}{*}{\begin{tabular}[c]{@{}l@{}}Post-covid\\ changes\end{tabular}} & \begin{tabular}[c]{@{}l@{}}Yes, I use a safer messaging \\ app now\end{tabular} & 20 & 20 & 0 & 40 \\
 & \begin{tabular}[c]{@{}l@{}}Yes, I didn’t use any \\ messaging app before, but \\ do use it now\end{tabular} & 20 & 23 & 1 & 44 \\
 & \begin{tabular}[c]{@{}l@{}}No, I did use it before and \\ use it now\end{tabular} & 114 & 47 & 4 & 165 \\
 & \begin{tabular}[c]{@{}l@{}}No, I didn’t use it before and \\ don’t use it now\end{tabular} & 18 & 6 & 0 & 24 \\
 & Other & 7 & 2 & 0 & 9 \\
Total &  & 179 & 98 & 5 & 282 \\ \hline
\end{tabular}
\vspace{0.8em}
\caption{Gender * post-covid changes crosstabulation}
\label{tab:gender*covid}
\end{table}
\begin{table}[h!tbp]
\centering
\resizebox{\textwidth}{!}{
\begin{tabular}{llcccccccc}
\hline
 & \multicolumn{1}{c}{} & \multicolumn{7}{c}{Disciplines} & Total \\ \cline{3-10} 
 & \multicolumn{1}{c}{} & \begin{tabular}[c]{@{}c@{}}Engi- \\ neering(f)\end{tabular} & \begin{tabular}[c]{@{}c@{}}Social \\ Sciences (f)\end{tabular} &  \begin{tabular}[c]{@{}c@{}}Sciences\\ (f)\end{tabular} & \begin{tabular}[c]{@{}c@{}}Bio-\\ medicine (f)\end{tabular} & \begin{tabular}[c]{@{}c@{}}Human-\\ ities(f)\end{tabular} & \begin{tabular}[c]{@{}c@{}}Technol-\\ ogy(f)\end{tabular} &  \begin{tabular}[c]{@{}c@{}}Other\\ (f)\end{tabular} & \\ \cline{2-10} 
\multirow{4}{*}{\begin{tabular}[c]{@{}l@{}}Post-covid\\ changes\end{tabular}} & \begin{tabular}[c]{@{}l@{}}Yes, I use a safer \\ messaging app now\end{tabular} & 7 & 11 & 5 & 5 & 1 & 10 & 1 & 40 \\
 & \begin{tabular}[c]{@{}l@{}}Yes, I didn’t use any \\ messaging app before, \\ but do use it now\end{tabular} & 13 & 12 & 5 &  4 &  2 &  7 & 1 & 44 \\
 & \begin{tabular}[c]{@{}l@{}}No, I did use it before \\ and use it now\end{tabular} &60 &  26 &  20 &  4 & 0 & 53 &  2 &  165\\
 & \begin{tabular}[c]{@{}l@{}}No, I didn’t use it \\ before and don’t \\ use it now\end{tabular} & 14 &  4 & 1 & 2 & 1 & 1 & 1 & 24 \\
 & Other & 5  & 0 & 1 & 1 & 1 & 1 & 0 & 9 \\
Total & & 99 &  53&  32&  16&  5&  72&  5&  282\\ \hline
\end{tabular}}
\vspace{0.8em}
\caption{Discipline * post-covid changes crosstabulation}
\label{tab:discipline*covid}
\end{table}
Finally, we took a closer look at the post-covid changes and the variables
gender, sector, and discipline for which the chi-square test is statistically
significant as shown in Figures \ref{figure:gender*postcovid_chisq},
\ref{figure:sector*postcovid_chisq}, \ref{figure:discipline*postcovid_chisq}
respectively. Each figure shows on the left, a graph that represents the
Pearson's residuals of the chi-square test results and a table on the right part
that shows the contribution of each cell to the test. In the Pearson's residual
graphs, \textcolor{black}{the responses to the question about the use of messaging apps were shortened for the sake of clarity according to}: 1)
\textit{Did/do} stands for \textit{Yes, I did use it before and do use it now};
2) \textit{Didn't/do} for \textit{Yes, I didn’t use any messaging app before,
  but do use it now}; 3) \textit{safer} for \textit{Yes, I use a safer messaging
  app now}; 4) \textit{Didn't/Don't} for \textit{No, I didn’t use it before and
  don’t use it now}.

As mentioned, the graphs on the left hand side represent the Pearson's
residuals, showing the difference between the observed and the expected values
for each cell. Thus, large residuals indicate that variables are not truly
independent. In our case, the blue shows positive contributions and red negative
contributions to the test. Moreover, the saturation shows how large the
contribution is in contrast to the expected value by chance. Additionally,
tables on the right side show adjusted standardized residuals that according to
\cite{agresti2013categorical}, if greater than +/-2 for cases with few cells,
indicate lack of fit of $H_0$ (in boldface in our table). Complementary, the
table also shows the percentage of contribution to the test of each cell
(highest percentage per column also in boldface).

Regarding gender in Fig. \ref{figure:gender*postcovid_chisq}, the results show a
high positive contribution from female educators that did not use a messaging
app before the COVID-19 pandemic but do use it now; approximately 31\% of the
test results are explained by this cell. In contrast, male educators that
respond the same were less than expected (negative contribution that explains
17\% of the test results). Also, it is important to highlight the female
educators that kept using these apps before and after the pandemic and
correspondingly the negative contribution of male educators to the same case
(less than expected by chance).

With respect to the sector, the results in Fig. \ref{figure:sector*postcovid_chisq} show a positive contribution from university educators being reluctant to use any messaging apps before or after the pandemic: the positive contribution of their responses explain 25\% of the test results with an adjusted residual of 2.94. On the contrary, vocational teachers were more open to it, showing a negative Pearson's residual. Also, we find larger values than expected in vocational teachers that kept using these apps after the pandemic. 

Finally, statistical significance was found for the discipline and post-covid
changes. Regarding disciplines (see
Fig. \ref{figure:discipline*postcovid_chisq}), the number of educators from
Humanities and Social Sciences that started using messaging apps after the
pandemic is larger than expected and these number of responses explains about
19\% of the chi-square test results. Also, the number of educators from
Technology or Engineering that started using safer alternatives \footnote{This addresses the issue of insecure/unsafe applications which do not encrypt messages or can be easily compromised exposing information} are less than
expected (compared to the educators from other disciplines). Bear in mind that
some disciplines were merged to avoid very low number of responses for some of
the cells.


\subsection{RQ2 - Which kind of chatbots would teachers find useful in their classes?}


For answering RQ2, teachers were provided with a list of different potential chatbot functionalities (use cases) and were requested to respond on whether each given use case would be useful in their classes. The findings are summarized in Table \ref{tab:chatbottypes}. 

\begin{table}[h!tbp]
\begin{tabular}{lllll}
\hline
\multicolumn{1}{c}{Chatbot use cases} & \multicolumn{2}{c}{Yes} & \multicolumn{2}{l}{No} \\ \cline{2-5} 
\multicolumn{1}{c}{} & \multicolumn{1}{c}{Frequency} & \multicolumn{1}{c}{\%} & Frequency & \multicolumn{1}{c}{\%} \\ \hline
Answering to students' FAQs & 148 & 52.5 & 134 & 47.5 \\
Assigning student grades & 113 & 40.1 & 169 & 59.9 \\
Facilitating agenda information & 171 & 60.6 & 111 & 39.4 \\
Sharing class materials & 136 & 48.2 & 146 & 51.8 \\
Others & 25 & 8.9 & 257 & 91.1 \\ \hline
\end{tabular}
\caption{Perceived useful chatbot use cases}
\label{tab:chatbottypes}
\end{table}

The most favourable use case for chatbots in
Higher Education and vocational training is their use for the facilitation of an
agenda formation (171 positive response, 60.6\%), followed by the FAQs use case
(148 positive responses, 52.5\%) and the sharing class material use case (136;
48.2\%). A chi-square test of independence was performed to examine the relation
among participants’ preferences for particular chatbot use cases. Out of the 171 teachers who consider useful the use of
chatbots for agenda preparation in the class, 103 also consider useful chatbots’
use for FAQs. The relation between agenda and the FAQs use case was significant, $X^{2}_{(1, N = 282)} =
10.467$, $p =0.001$. The frequencies cross tabulated are given in Table
\ref{tab:agenda*faq}. 

\begin{table}[h!tbp]
\begin{tabular}{llccc}
\hline
 & \multicolumn{1}{c}{} & \multicolumn{2}{c}{FAQs} & Total \\ \cline{3-5} 
 & \multicolumn{1}{c}{} & Yes & No &  \\ \cline{2-5} 
\multirow{2}{*}{Agenda} & Yes & 103 & 68 & 171 \\
 & No & 45 & 66 & 111 \\
Total &  & 148 & 134 & 282 \\ \hline
\end{tabular}
\vspace{0.8em}
\caption{Agenda use case * FAQs use case crosstabulation }
\label{tab:agenda*faq}
\end{table}

\begin{figure}[h!tbp]
\begin{center}
    \begin{minipage}[t]{0.85\textwidth}
        \centering
\begin{knitrout}
\definecolor{shadecolor}{rgb}{0.969, 0.969, 0.969}\color{fgcolor}
\includegraphics[width=\maxwidth]{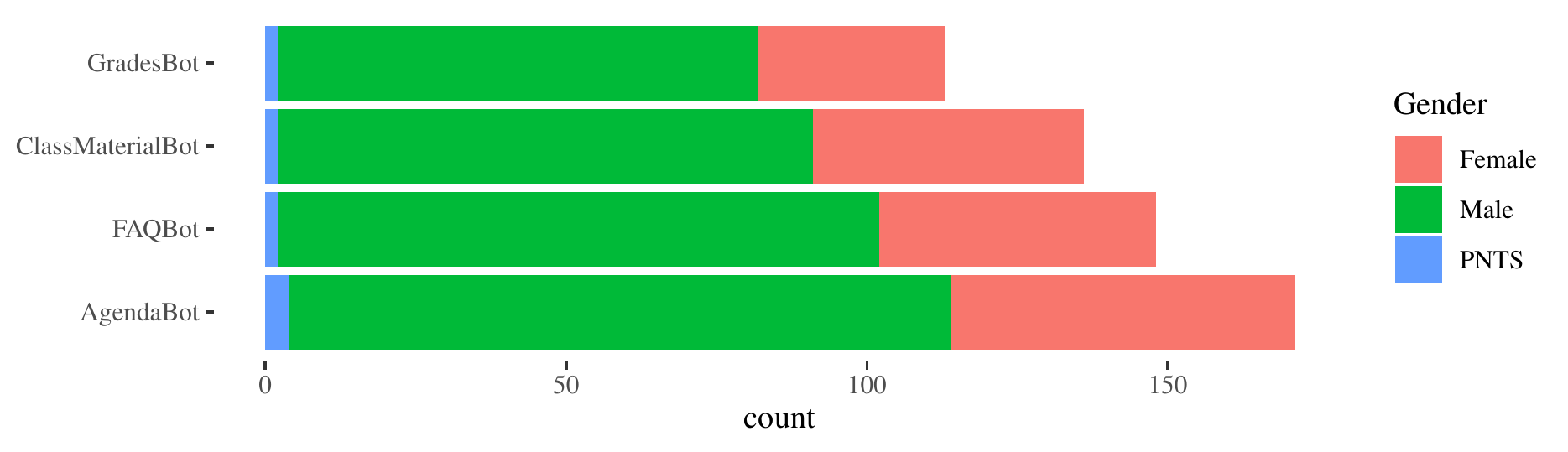} 

\end{knitrout}
        \vspace{-0.5cm}
        \caption{Count of types of chatbots for class perceived as the most useful for teachers grouped by gender (PNTS stands for Prefer Not To Say).}
        \label{figure:gender*chatbots_types}
    \end{minipage}
    \begin{minipage}[t]{0.85\textwidth}
        \centering
\begin{knitrout}
\definecolor{shadecolor}{rgb}{0.969, 0.969, 0.969}\color{fgcolor}
\includegraphics[width=\maxwidth]{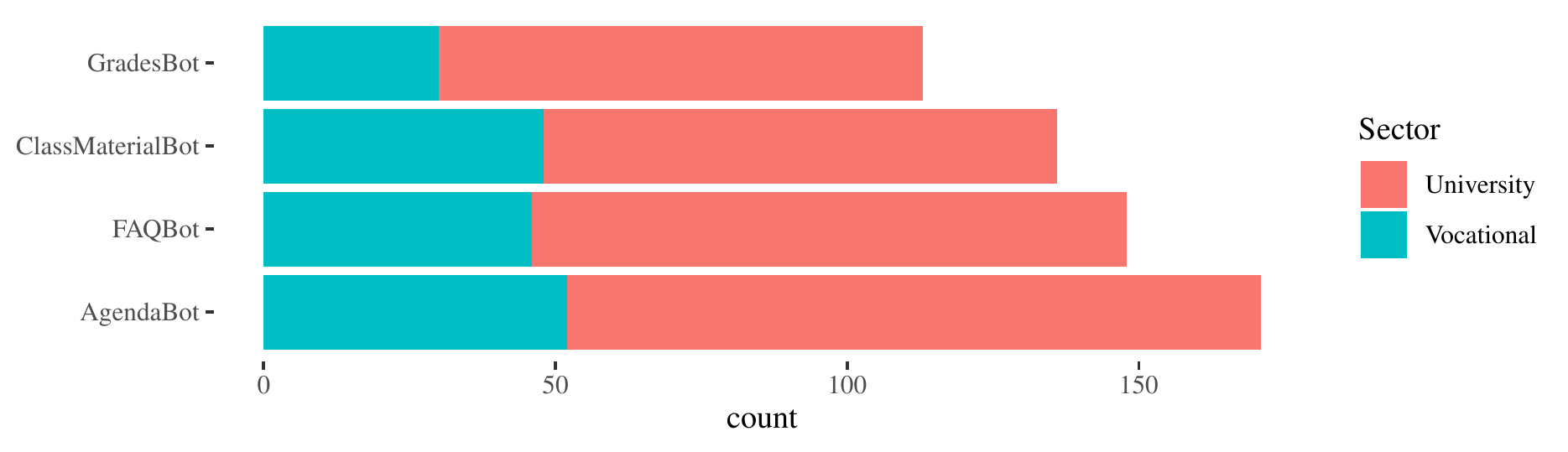} 

\end{knitrout}
        \vspace{-0.5cm}
        \caption{Count of types of chatbots for class perceived as the most useful for teachers grouped by sector.}
        \label{figure:sector*chatbots_types}
    \end{minipage}
    \begin{minipage}[t]{0.85\textwidth}
        \centering
\begin{knitrout}
\definecolor{shadecolor}{rgb}{0.969, 0.969, 0.969}\color{fgcolor}
\includegraphics[width=\maxwidth]{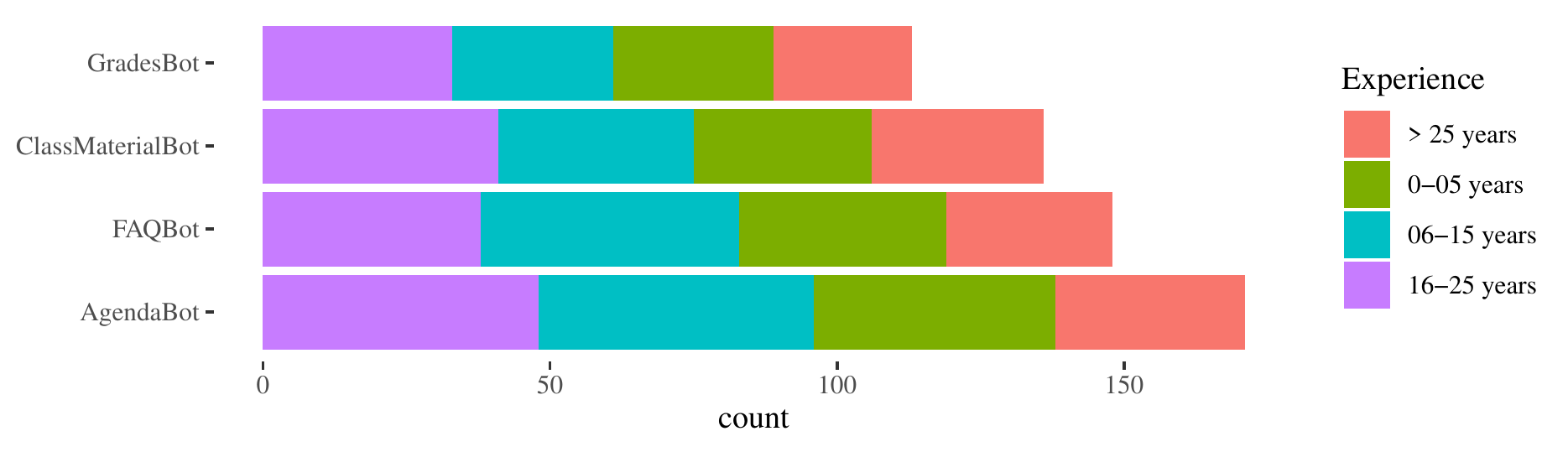} 

\end{knitrout}
        \vspace{-0.5cm}
        \caption{Count of types of chatbots for class perceived as the most useful for teachers grouped by experience.}
        \label{figure:experience*chatbots_types}
    \end{minipage}
    \begin{minipage}[t]{0.85\textwidth}
        \centering
\begin{knitrout}
\definecolor{shadecolor}{rgb}{0.969, 0.969, 0.969}\color{fgcolor}
\includegraphics[width=\maxwidth]{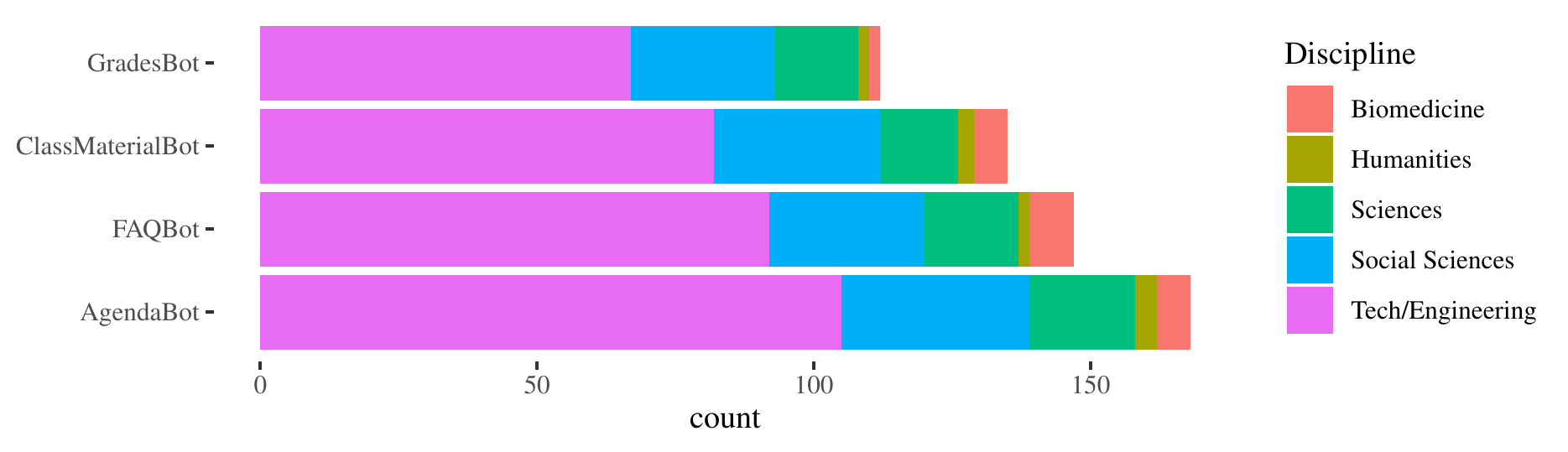} 

\end{knitrout}
        \vspace{-0.5cm}
        \caption{Count of types of chatbots for class perceived as the most useful for teachers grouped by discipline.}
        \label{figure:disc*chatbots_types}
    \end{minipage}
\end{center}
\end{figure}

Answers to these questions are plotted in
Figs. \ref{figure:gender*chatbots_types}, \ref{figure:sector*chatbots_types},
\ref{figure:experience*chatbots_types}, \ref{figure:disc*chatbots_types} grouped
by gender, sector, years of experience in education, and discipline
respectively. As it was a multiple choice question, the counts are over the
total number of people that answered.

\subsection{RQ3 - Which kind of interaction do teachers prefer with their students?}


\begin{figure}[h!tbp]
\centering
\begin{knitrout}
\definecolor{shadecolor}{rgb}{0.969, 0.969, 0.969}\color{fgcolor}
\includegraphics[width=\maxwidth]{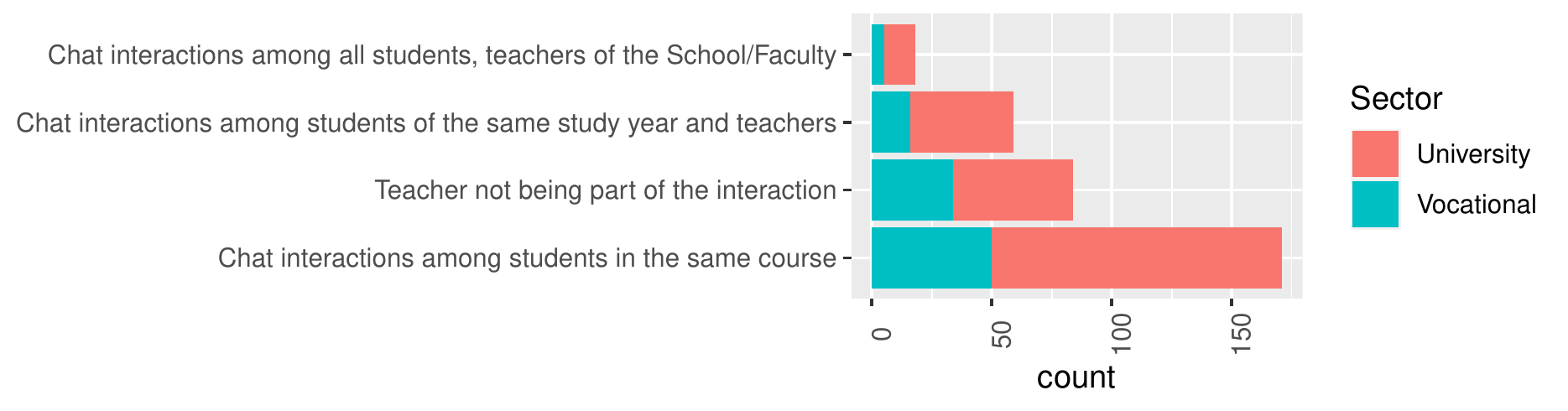} 

\end{knitrout}
\caption{Distribution of teachers' preferences for the chat groups with their students: from groups only with their students from a specific course to groups with greater social interaction with all students in their School or Faculty.}
\label{figure:chatbots_messagingorganization}
\end{figure}

Since chatbots are intended to mediate or help in this interaction, it is
essential to understand the kind of interaction teachers prefer. Chatbots should
address those modes, and not others. In order to find out these modes, and thus
answering RQ3, teachers were provided with a list of different kind of
interactions that may take place among students and between students and the
teacher, with the use of messaging apps. The findings are summarized in Table
\ref{tab:kindinteractions}.

\begin{table}[h!tbp]
\begin{tabular}{lllll}
\hline
\multicolumn{1}{c}{Kind of interactions} & \multicolumn{2}{c}{Yes} & \multicolumn{2}{l}{No} \\ \cline{2-5} 
\multicolumn{1}{c}{} & \multicolumn{1}{c}{Frequency} & \multicolumn{1}{c}{\%} & Frequency & \multicolumn{1}{c}{\%} \\ \hline
\begin{tabular}[c]{@{}l@{}}Chat interactions among \\ students in the same course\end{tabular} & 166 & 59.1 & 115& 40.9 \\
\begin{tabular}[c]{@{}l@{}}Chat interactions among students, \\ teachers of the School/Faculty\end{tabular} & 19 & 6.7 & 266 & 94.3 \\
\begin{tabular}[c]{@{}l@{}}Chat interactions among students \\ of the same study year and teachers\end{tabular} & 58 & 20.5 & 224 & 79.4 \\
\begin{tabular}[c]{@{}l@{}}Teacher not being part of the \\ interaction\end{tabular} & 81 & 28.7 & 201 & 71.3 \\ \hline
\end{tabular}
\vspace{0.8em}
\caption{Kind of interactions preferred}
\label{tab:kindinteractions}
\end{table}

As it can be seen by the answer to the first and last question, in general
teachers do not want to participate in a chat group with students; either they
want to simply leave the students alone in their own chat group, or otherwise
they prefer not to be part of that interaction. In general, that is going to be
the case no matter what; it is well known that students organize their own chat
groups with many (and not always conveyable) intentions, so teachers do not want
to take any part in these informal or non institutionally-supported chat
groups. Overwhelmingly, they do not want to participate in these kind of chat
groups with students, but even less so if it includes the rest of the faculty.

The answer, then, to this research question, is that, in general, teachers do
{\em not} want to have interaction with students in a chat group. To a certain
point, this would seem to contradict results \cite{gachago2015crossing}, although this might be due to cultural
attitudes, or other factors such as the average size of classes. It does
confirm, however, that the challenges cited in that study, and possibly others,
are an obstacle to the adoption of mobile (and other) instant messaging among
the community of surveyed teachers.

Regarding the social factor of chat groups that teachers use in class, according
to Fig. \ref{figure:chatbots_messagingorganization} the vast majority prefer
small groups only with the students from the same course. These are more focused
groups with specific goals and dedicated to the organization of the course and
its tasks, and from the pedagogic point of view \textcolor{black}{it also seems more appropriate to improve
the learning process}. Interestingly, and as a cross-check of the answers above,
about 30\% of teachers consider they should not be part of the chat group. This
might seem to contradict the results of the other survey, but in fact, being a
result of different surveys, to a certain point affirms the same thing: there is
a great amount of teachers that would be against being in a chat group with
students. \textcolor{black}{However, students will still need to get the services
that the university, through chatbots, provide, thus opening the door for deployment of chatbots
without the intervention of the professor, simply tapping university provided
services \cite{bernier2002services}.}
Whether that's a slim majority or not, that's open to debate (and
would probably need a more pointed survey focused on this), but what is clear is
that forcing teachers to create chat groups with students and participate in
them would create a certain amount of resistance. Also, only university teachers
find interesting a group with all the students and teachers in their own Faculty
or School. The lack of teachers from vocational education here may be the
consequence of using specific language such as ``Faculty''. Moreover, the fact
that many universities and schools already use these groups for administrative
and social interaction (e.g. \cite{dibitonto2018chatbot}) might be the reason
for the low percentage of teachers that selected this response.





\begin{figure}[h!tbp]
\begin{center}
    \begin{minipage}[t]{0.3\textwidth}
        \centering
\begin{knitrout}
\definecolor{shadecolor}{rgb}{0.969, 0.969, 0.969}\color{fgcolor}
\includegraphics[width=\maxwidth]{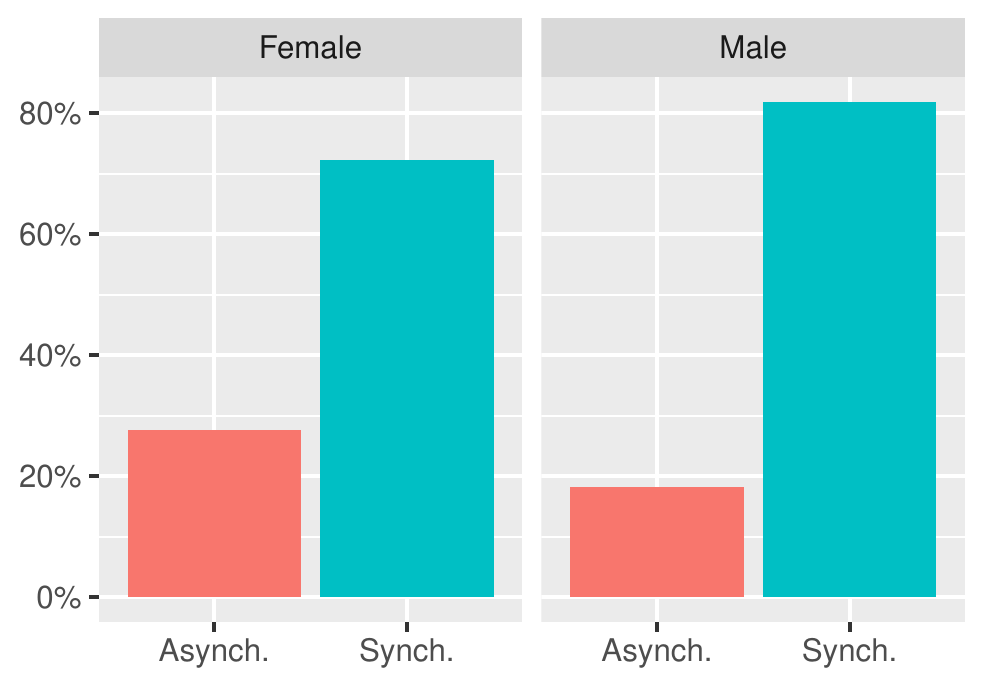} 

\end{knitrout}
    \end{minipage}
    \begin{minipage}[t]{0.3\textwidth}
        \centering
\begin{knitrout}
\definecolor{shadecolor}{rgb}{0.969, 0.969, 0.969}\color{fgcolor}
\includegraphics[width=\maxwidth]{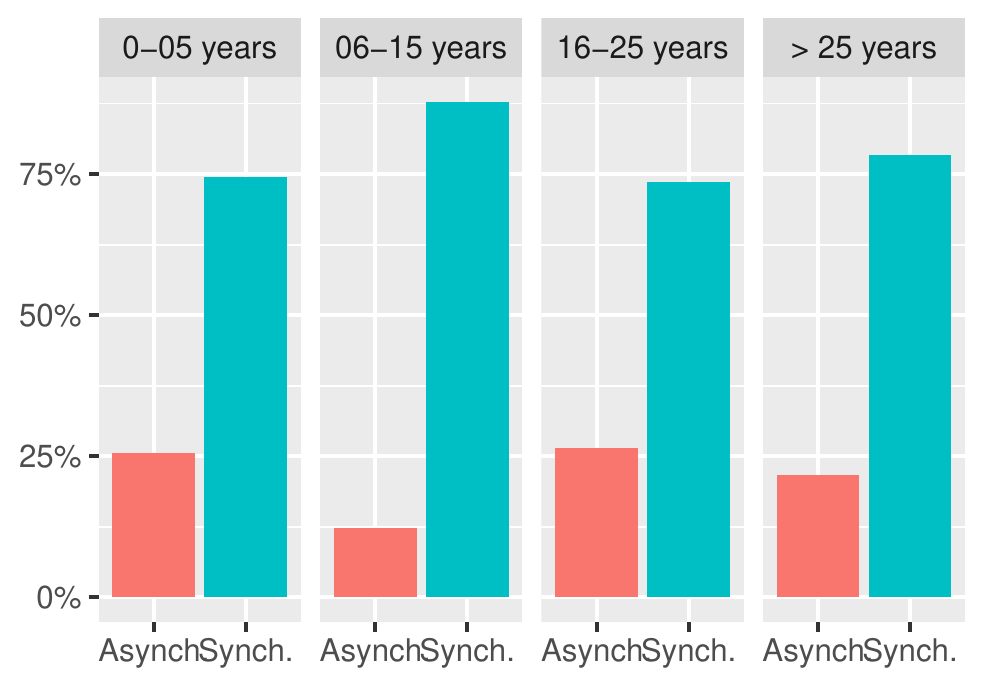} 

\end{knitrout}
    \end{minipage}
    \begin{minipage}[t]{0.6\textwidth}
        \centering
\begin{knitrout}
\definecolor{shadecolor}{rgb}{0.969, 0.969, 0.969}\color{fgcolor}
\includegraphics[width=\maxwidth]{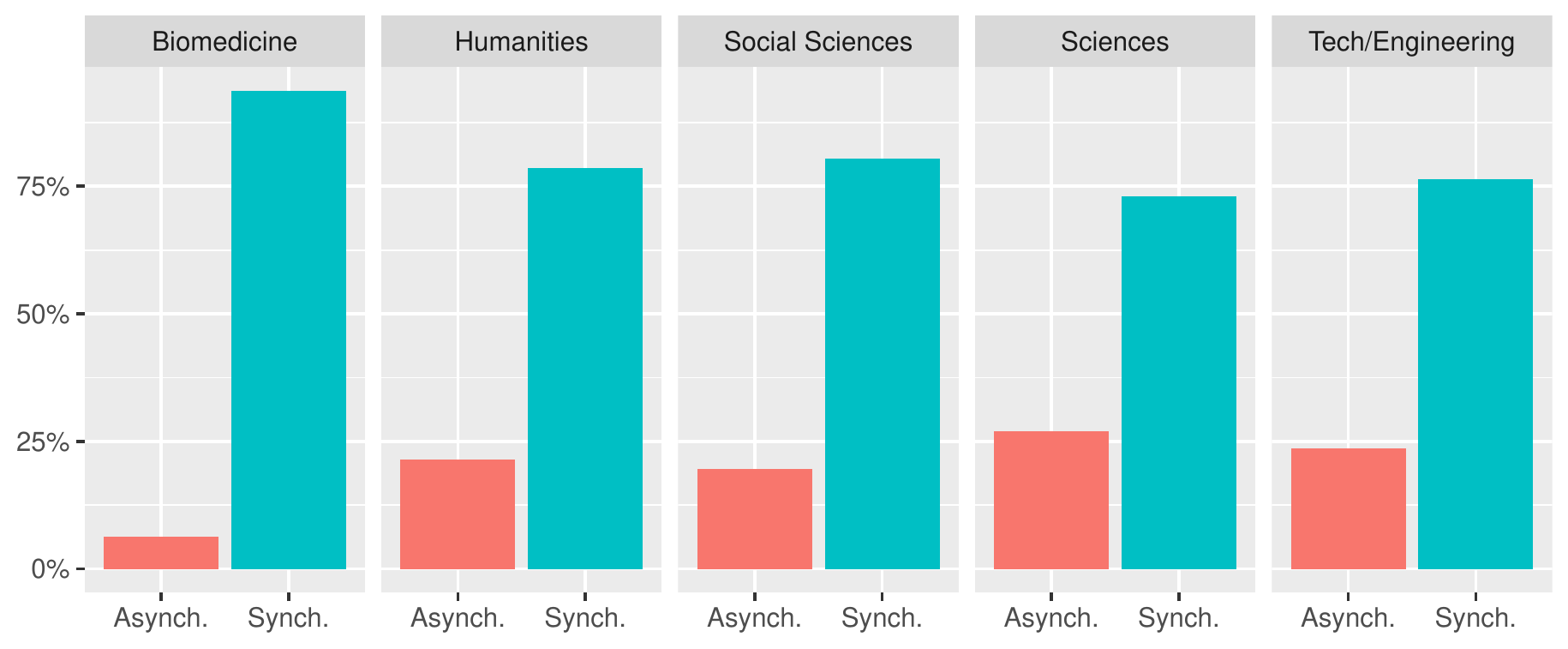} 

\end{knitrout}
    \end{minipage}
\end{center}
    \caption{Results of teachers' communication preferences: synchronous or asynchronous interaction with students (grouped by gender, years of experience, and discipline).}
\label{figure:chatbots_synch}
\end{figure}


\subsection{RQ4 - What kind of interaction media features do teachers value the most? }



\begin{table}[h!]
\begin{tabular}{lcccc}
\hline
Interaction features & \multicolumn{2}{c}{Yes} & \multicolumn{2}{c}{No} \\
\hline
                    &  Freq. & \% &  Freq. & \% \\
                    \cline{2-5}
                
Analytics & 108 & 52.7 & 97 & 47.3 \\
Connectivity & 119 & 58.0 & 86 & 42.0 \\
Familiarity & 121 & 59.0 & 84 & 41.0 \\
Hidden Phone Number & 113 & 55.1 & 92 & 44.9 \\
Horizontally & 134 & 65.4 & 71 & 34.6 \\
Official formation & 42 & 20.5 & 163 & 79.5 \\
Pluggability & 65 & 31.7 & 140 & 68.3 \\
Sustainability & 157 & 76.6 & 48 & 23.4\\
Unidirectionality & 27 & 13.2 & 178 & 86.8 \\
Officiality & 150 & 73.2 & 55 & 26.8 \\
Synchrony & 45 & 22.0 & 160 & 78.0\\
Other & 5 & 2.4 & 200 & 97.6 \\
\hline
\end{tabular}
\vspace{0.8em}
\caption{Interaction media features valued by teachers. Details about each of these features can be read in Appendix I (Section \ref{app:questions}, second survey).}
\label{tab:interaction}
\end{table}

\textcolor{black}{As it can be seen in the table, the bulk of responders express their wish to use a sustainable and official application, i.e. both being approved or provided by the University (or the educational centre), and also maintained by the technical staff of the centre instead of giving this responsability/task to the teacher.
In addition, it is very important for the teachers that the used tools that all the members of the class communicate, including themselves; as well as being already known or used in everyday tasks, e.g. Telegram or WhatsApp.}

A chi-square test of independence was performed to examine the potential relationship among the various features. The statistically significant correlations are given in Table \ref{tab:interactionfeatures}.  

\begin{table}[h!tbp]
\centering
\resizebox{\textwidth}{!}{\begin{tabular}{lllllllll}
\hline
 & Connectivity & Familiarity  & Horizontality & \begin{tabular}[c]{@{}l@{}}Official\\  formation\end{tabular} & Pluggability & Sustainability & Unidirectionality & Officiality \\ \hline
Analytics &  $12.172^{***}$ & $4.855^{*}$ & ns & $4.144^{*}$ &  $12.490^{***}$ & $4.314^{*}$ & ns & ns \\
Connectivity & - & ns & $5.971^{***}$ & ns & $16.286^{***}$ & ns & ns & ns \\
Hidden phone &  &  & ns & $5.678^{*}$ & $4.683^{*}$ & ns & $4.515^{*}$ & ns \\
Horizontality &  &  - & ns & ns & ns & ns & ns & $5.453^{*}$ \\
\begin{tabular}[c]{@{}l@{}}Official\\  formation\end{tabular} &  & &  & - & $8.163^{**}$ & ns & ns & ns  \\
\end{tabular}}
\vspace{0.8em}
\caption{Interaction media features analysis ($^{***}p<0.001$, $^{**}p<0.01$, $^{*}p<0.05$). Columns and rows with no significant interaction have been suppressed for clarity. Every cell value represents $X^{2}_{(1, N = 205)}$. Details about each feature can be read in Appendix I.}
\label{tab:interactionfeatures}
\end{table}

\begin{figure}[h!tbp]
\begin{center}
\includegraphics[width=\textwidth]{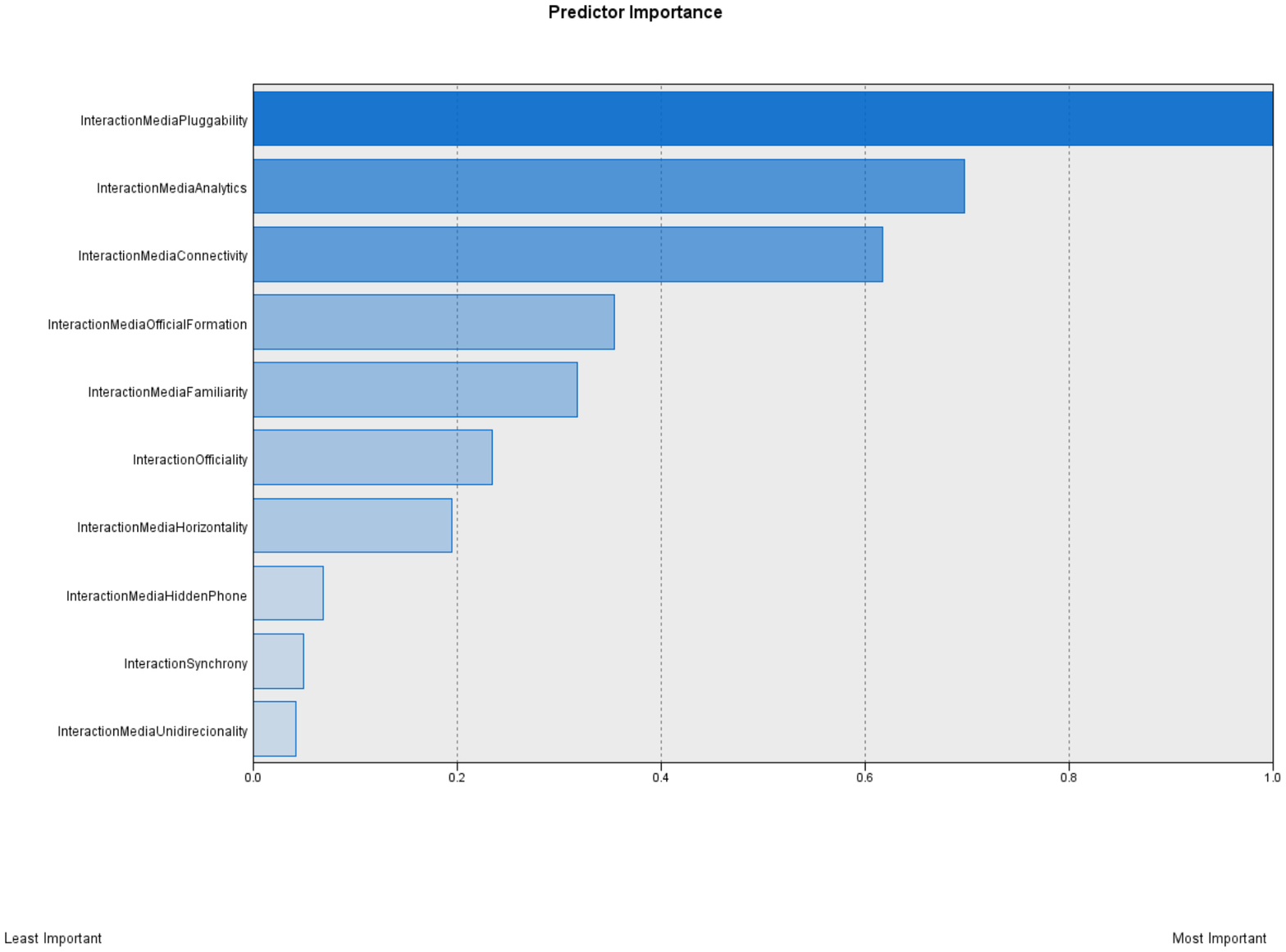}
\end{center}
    \caption{Two-step cluster analysis for interaction media features: predictor importance}
\label{figure:predictor}
\end{figure}

Finally, a two-step cluster analysis was conducted, with a log-likelihood
distance measure adopted, to explore how the use cases could be grouped based on
their preferences of specific interaction media features, and which features
have larger predictor importance for the clustering (see
Fig. \ref{figure:predictor}). Two clusters occurred from the analysis: cluster 1
(n=78, 60.9\%) is formed with instructors who did not value interaction media
pluggability, connectivity, official formation, but valued interaction media
analytics, familiarity, support by the institution; cluster 2 (n=50,
39.1\%) groups instructors that did not value media
analytics, official formation, familiarity, but valued pluggability,
connectivity, and support by the institution. Interaction media pluggability
appears to have the most important predictor importance in the clustering of
cases (predictor importance = 1.0), whilst media unidirectionality was the least
important factor (predictor importance = 0.04). The cluster quality is fair, but
not a good one (silhouette measure of cohesion and separation).

\section{Discussion}

Our initial intention in the design of these surveys was to probe the opinions
of tertiary education teachers in the introduction of chatbot technologies in
class. We wanted to find the answer to four research questions, of which the
first was to check whether teachers already used some kind of messaging
application, and the one they preferred. Depending on the answer, different
technologies could be (or not) available for the introduction of chatbots. In
this case, the design of the survey introduced some ambiguity on what we really
consider ``messaging technology'', that was eventually resolved by the answers;
apparently, anything that sends messages (even if they are not instant, even if
they do not have a dedicated application) was the concept in the mind of the
surveyed.  \textcolor{black}{In this case, we would like 
to add a clarification. While we had in mind, as indicated by the possible answers, 
instant messaging applications when we elaborated this question,
we included these institution-provided applications mainly for completion. But
it should be noted that, in most cases, they are not {\em instant} messaging
apps in the way Telegram or Facebook Messenger are; in most cases they are
messaging facilities provided by learning management systems such as (usually)
Moodle. So the answer to this question must be understood in two different ways:
their preference for the {\em provider} of the technology (the university itself, any
company) as well as the {\em type} of technology (instant, app based messaging
systems vs. web-based messaging systems integrated in another learning management system).}

This inclusion of university-provided messaging applications in
the survey might explain why there are so few, just 6.7\%, that answer they do not 
use any messaging application. In general, universities provide official channels
to communicate with students official academic information such as the exam schedule
or the grades. The statistical mode is to use that kind of messaging. Some other
people might not use it, but use some extraofficial or opt-in channels such as the other included in the survey; at the end, between these official and extraofficial channels most people cover their needs to communicate with students. This matches
our subjective experience, in general.

Additionally, the question on COVID reflected the big challenge that represents
changing habits, even in the case of a major crisis. Very few responders actually
changed their habits, although comparing women vs. men as well as those teaching social
sciences show some significant difference. We do not have any kind of explanation for
this behaviour, other than technologies are perceived, and adopted, in a
different way depending on background and, apparently, gender.

One of the most interesting questions that can arise from the analysis of the
results, and that should require further study, is: Are teachers really aware
of the possibilities of chatbots in the classroom? The survey 
only listed a few functionalities for the chatbots and just a few subjects selected 
the option "Others". This may suggest  that most teachers did not really understand 
at this point the actual potential of chatbots, since the responses listed were by no means exhaustive, and if the responders were more acquainted with the possibilities of the technology, possibly this response would have shown up many more times. Another way to look about it is 
that, unless you provide the specific instructions on how to use a
tool, the teaching staff is already too busy or burn out to think about
innovation.

We can speculate that their opinions towards chatbots are informed by negative
experiences with customer support chatbots, which is what most people will have
experienced. This will probably require further study, but also will need to be
taken into account when introducing chatbot technologies in the classroom: 
it will have to be as simple, and as satisfactory, as possible, 
to (possibly) change this existing perception.

Regarding gender grouping, both genders agree in their uses, so it seems not
relevant at all. When looking at the plot containing the sector differentiation
it is interesting to observe how university teachers prefer in a higher
proportion to use them for grades and FAQs in comparison with other tertiary education teachers. 
This seems reasonable as the frequency of seeing the students
is lower at the University so, an automatic tool for grade notification for
continuous evaluation seems more helpful. 
Furthermore, the high student/teacher
ratios that university teachers have to deal with could be another reason to
have a higher number of them willing to use chatbots for FAQs. This could
ameliorate the number of misunderstandings and email overload that so many
students might generate.

The results, when grouping by experience, confirm that
experienced teachers are not afraid of changes and are willing to accept new
technological challenges; at least, in the same way that younger teachers
are. From this concrete plot it is interesting as well to observe the fact that very few
young/inexperience teachers are willing to use chatbots for other uses. This
reinforces the hypothesis thrown at the beginning of this section that younger
teachers have less experience to consider new ways of using chatbots in their daily work.

As it appears in the responses presented above, teachers’ responses are equally
distributed between positive and negative in relation to the media features:
analytics, connectivity, and hidden phone number; that is, none of them is a
prevailing factor on choosing one technology over another. However, for the features: familiarity,
horizontally, sustainability and officiality the majority provided a positive
response, designating their preference to these features. On the other hand, the
majority of teachers seems to consider not valuable the presence of official
formation, pluggability, unidirectionality and synchrony in the interaction
media features of messaging apps.

From a qualitative point of view, these results show consensus in a few key aspects:
\begin{itemize}
 \item Teachers want to have support from their corresponding institution. This
   is probably due to the strict European data regulations \textcolor{black}{on data protection},
   but also to avoid the overhead required to \textcolor{black}{sign up new students every year.
   In general support from specific formation and access to IT help desk for specific features
   enhances the possibility of adoption by the teachers or the possibility of a successful
   adoption that improves the learning outcomes. This is supported by the result of the two
   surveys: they prefer whatever "messaging" application is provided by the university, and
   also they value sustainability as well as "officiality" of the platform, as shown in
   Table \ref{tab:interaction}}.
 \item Teachers need sustainability. The main reason is that implementing
   changes is expensive in terms of effort and prone to errors in the first
   years of implementing new procedures, therefore, it is logical that if
   chatbots or messaging platforms are to be introduced, they want this change
   to be as permanent as possible.
 \item Keep it Simple. Media interaction requirements are simple and only a 2.4\% is requesting more features. This could be a sign of technological burn out produced by the previous and current courses where, due to the pandemic, the use of computer has increased significantly.
\end{itemize}

An analysis of survey questions related to RQ4 seems to indicate that teachers
prefer to share an interaction space with their student, as 
shown in Tables \ref{tab:kindinteractions} and \ref{tab:interaction}, where a majority of responders
indicate they {\em don't} want unidirectional communication; a preference of
{\em horizontal} interaction, but also the answer to the kind of
chatbots preferred that features FAQ bots prominently, which might indicate a
need to delegate or offload part
of the burden of answering, and attending, every single question posed by
students. In an horizontal setting, other students can answer if the teacher
does not do it immediately. This might also be the reason why the possibility of
using tools from which analytics can be extracted is also valued by teachers:
that way, activity in the chat tool, answering questions, regularity, can all be
observed and valued as part of the process of assessing achievement of learning
objectives by students.


\section{Conclusions}
\label{sec:concl}

The key findings of this study shed the light on educators’ preference to use
messaging applications that their institutions support.  Technology adoption
literature often \textcolor{black}{focuses} on users’ perceptions of the technology’s usefulness and
ease of use as important prerequisites for successful adoption and
utilization. Nevertheless, in higher education, institutions’ role in
integrating these tools to their educational systems can improve the uptake of
these applications and shape the social and educational experiences of their
students. \textcolor{black}{To accomplish this fruitful integration}, institutions have to ensure that
these messaging applications not only are GDPR (General Data Protection Regulation) compliant to keep students’ data
secure, but also should provide IT support to all stakeholders who use the
applications.

Comparing these results with those obtained in student surveys
in \cite{MoraChatbots2021}, we have understood better the differences between the teachers’
points of view and/or intentions and that of their students when using the
messaging applications in higher education.  Teachers tend to adopt technologies
supported by their institutions.  The fact they do so could be caused by a desire to ensure that their
universities can oversee their efforts in supporting their students during the
learning process.  Another reason could be their familiarity with the technology
provided by their institutions. On the other hand, students use the
non-institution messaging applications to form informal discussion groups with
their peers.  It is worth noting that peer support and collaboration is
inseparable from learning \cite{timmis2012constant} and correlates positively
with higher retention in higher education \cite{o2014front}.  Therefore, both
perspectives are complementary and play different roles in promoting the
learning process.

As in many other methodologies that try to assess technology acceptance, answers to the survey suggest
that the introduction of simple, and institution-supported, instant messaging
and chatbot technologies would increase the perceived usefulness (which is one
of the key metrics in technology acceptance models). Since most institutions
already have some {\em virtual campus} or learning management system, adding
some easy automation, or connection to personal instant messaging tools, could
really help onboard the learning community on these new technologies.

What we are going to propose next is a possible process of
technology introduction that is compatible with the conclusions of this study,
but that would have to be piloted in order to check its value, and its
relationship with better learning outcomes, as well as higher teacher
satisfaction.
Once that initial introduction of institutional messaging
automation tools is done, teachers (and students as well) will
probably prefer the kind of bots that alleviate bureaucratic or repetitive
tasks, such as answering frequent questions or answers on class or assignment
deadlines, as indicated by their answers to the respective
questions. These will help the introduction of more complex chatbots that will
affect more directly learning outcomes, such as chatbots that help students
integrate in the class, or are able to identify (and address) learning problems
in students or in groups of them. These should also be accompanied by
analytics on student interaction, as well as possibly some natural language
processing (in vernacular language) that will help assess the general mood of
the class, and how different material (or external factors) affect it.

This introduction of chatbots and chatbot technology will be helped by the fact
that no discernible differences are found between different groups. Even if
chatbots do have some potential for personalization or customization, based on
the class material, student and teaching staff, the general introduction of the
technology can be done in a general way and for all disciplines, teaching
experience and gender.

In our survey, we could not find an answer about the right moment introducing chatbot technologies. Response to questions related to change of
behaviour during and after the first, stay at home, stage of the pandemic do not suggest that major (or minor)
crisis could be an opportunity to introduce new technologies, since it does not
bring major changes in attitude. A minor crisis would be, for instance, rollout
of a new higher education law\footnote{We seem to have one of these, at least in
  Spain, every 10 years or so.}, or introduction of new degrees. Although
external changes do offer the chance of piloting new technologies, they do not
seem to bring changes in attitudes in the teaching staff (which, after all, is
bound to be the same). In absence of a clear answer in this direction, the right
moment to introduce new technologies is always when the IT and managing staff is
ready to support it (since, as we have seen before, "official" support is one of
the factors that is most valued by teachers).

At any rate, the general feeling that transpires from the survey is that it is
essential for any institution to take into account the shareholder's opinion
when introducing chatbots. This is true almost across the board, but in the case
of chatbots (and instant messaging applications) their immediacy and the fact
that they can enter into what we could call the private sphere makes this even
more necessary.

One of the questions we made in the {\bf second survey}, related to
the messaging applications used by tertiary education teachers, opens a new line
of inquiry about what they {\em perceive} as such, and about how it is used.
Namely, the responses indicated that the teachers perceived as messaging
application not only what is usually called a chat or instant messaging app such
as WhatsApp or Microsoft Teams, but also the means provided by the university for
communicating with the student, such as a feature of grading applications that
will email the grade to the student. This implies that there is a need by
teachers, communication with students, also mostly unidirectional, and that it
does not matter so much how that need is covered. However, this will need careful
consideration, including how it ties with the automation of the learning
experience that the chatbots bring.

There are several future lines of work informed by our results of this
survey. For instance, the rollout of extensive videoconferencing and virtual
teaching solutions that the COVID pandemic has brought has also taught us a
series of lessons. It increases isolation, for instance, and decreases the
amount of synchronous contact that happens in the fringes of the classroom:
teaching staff offices, before and after class. A future line of work would be
focused on exactly this, and what kind of needs could be covered by chatbot
technology. We will create a series of international surveys that will
investigate this.

Finally, the full extent of chatbot technology is not really being examined in
these surveys. They can be connected to natural language processing engines with
sentiment analysis as well as other analytics. Examining the mood of the class,
as response to new material, assignment, exams or external events might help
will help any student-centred teaching strategy, which will help students (and
teachers) reach their learning objectives. This line of work, however, will be
based on the introduction of some pilot study, combined with initial opinion
assessment on the students.

\bibliographystyle{apalike}
\bibliography{edubots}

\begin{backmatter}


\section{Declarations}
\label{sec:decl}

\subsection{Availability of data and materials}

The datasets generated and analysed during the current study are available in the {\tt edubots-paper} repository, accessible from https://anonymous.url

\subsection{Competing interests}

The authors declare that they have no competing interests.

\subsection{Funding}

This work has been supported by EDUBOTS project, funded under the scheme Erasmus + KA2: Cooperation for innovation and the exchange of good practices - Knowledge Alliances (grant agreement no: 612446).

\subsection{Authors' contributions}

All the authors have contributed to the study presented in this manuscript. (Rest hidden for double blind)

JJM has been the leader, drafted the first versions of the survey and supervised the writing of the manuscript.

PAC has revised the paper and researched the state of the art, as well as contributed to the survey design.
AMM has revised the paper and researched the state of the art, as well as contributed to the survey design.
FB has made the analysis and processed data and contributed to the survey design.
NA has supervised the state of the art, and written the bulk of it, as well as contributed to discussion and conclusions.
OT has performed data analysis, and contributed to discussion and conclusions.
AG has contributed to the design of the survey, analysis, discussion and conclusions.

All authors have read and approved the final manuscript.

\subsection{Acknowledgements}

The authors want to thank to the educators who participated in the survey, and also people who have distributed the survey in their mailing lists, especially Mario García Valdez, Jesús Moreno León and the late José Raúl Canay.
We would also like to thank the anonymous reviewers that have contributed greatly to the improvement of this paper.

\section{Appendix I: Survey questions}
\label{app:questions}

The surveys have the same common initial demographic questions:\begin{itemize}
\item Degrees or titles where the teacher was working, either university or other tertiary education.
\item Discipline: Engineering, Social Sciences, Health and Bio, Sciences, Humanities, Other.
\item Gender: Man, Woman, Rather not say.
\item Age: 25-35, 35-45, 45-55, more than 55.
\item Experience on the job: up to 5 years, 5-15 years, 15-25 years, more than 25 years.
\textcolor{black}{
\item Type of degree: university, other tertiary education. This question was implicit in the first survey: there were two different versions of the survey. However, it was an explicit question in the second survey.}
\end{itemize}

\begin{figure}[h!tbp]
\begin{center}
\includegraphics[width=0.85\textwidth]{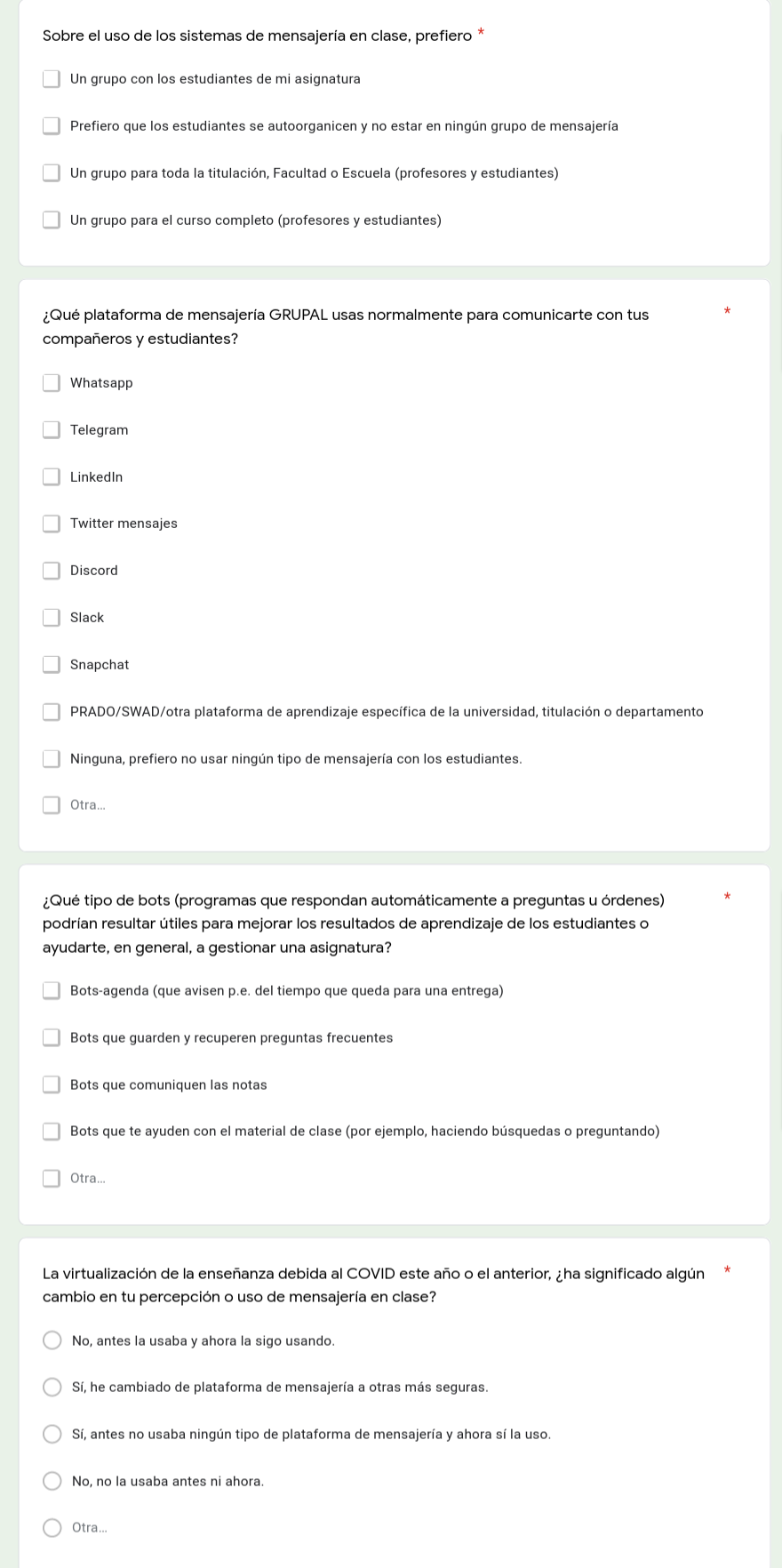}
\end{center}
    \caption{Screen capture of the Spanish-language form used for the \bf{first survey}}
\label{figure:survey1}
\end{figure}
The {\bf first survey}, shown in its original form in Figure \ref{figure:survey1}, which \textcolor{black}{focused on} the use of messaging apps in class, includes the following questions:

\textit{1. Regarding the use of messaging apps in class, you prefer ...}
\begin{itemize}
\item A chat group with the students in the same course
\item Students self-organize and with the teacher not being part of any chat group
\textcolor{black}{
\item A chat group with all the students and teachers of the School/High-School/Faculty
}
\item A chat group with all the students and teachers of the same study year
\end{itemize}

\textit{2. Which messaging app platform do you use to communicate with your peers or students?}
\begin{itemize}
\item WhatsApp
\item Telegram
\item LinkedIn
\item Twitter messages
\item Discord
\item Slack
\item Snapchat
\item The platform provided by your own institution
\item None, I do not use any messaging app to communicate with my students
\end{itemize}

\textit{3. Which kind of chatbots (software that automatically respond to questions or commands) could be useful to improve the learning results with your students or help you managing your course?}
\begin{itemize}
\item Agenda bots that e.g. remind students about project deadlines
\item Bots that collect and provide answers to frequent questions
\item Bots that inform students about their grades
\item Bots that help with the class materials (e.g. searching for topics about a concept, or asking about them)
\end{itemize}

\textit{4. The virtualization of teaching due to the COVID pandemic during this year and the previous one, did it mean any change in your viewpoint or use of messaging apps in class?}
\begin{itemize}
\item No, I did use it before and use it now
\item Yes, I started using safer alternatives
\item Yes, I did not use any messaging platform before but I do use it now
\item No, I did not use any messaging platform before or now
\end{itemize}

\begin{figure}[h!tbp]
\begin{center}
\includegraphics[width=\textwidth]{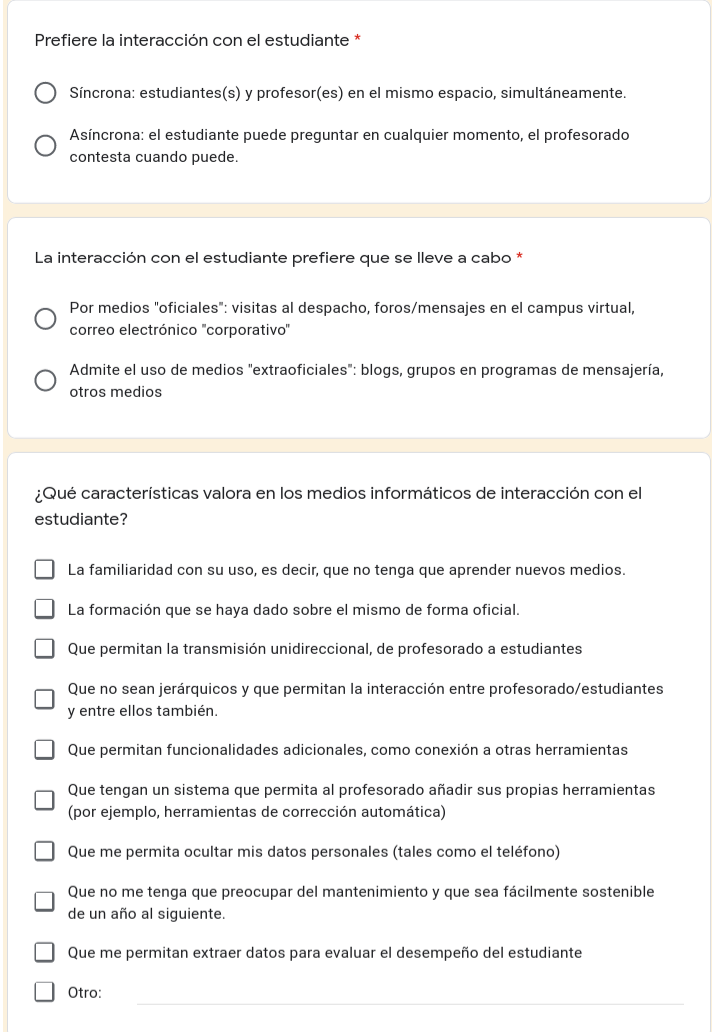}
\end{center}
    \caption{Screen capture of the Spanish-language form used for the \bf{second survey}}
\label{figure:survey2}
\end{figure}
The {\bf second survey}, shown in its original form in Figure \ref{figure:survey2}, followed the same paths, only in this case there was a single questionnaire, and with a different scope, focusing on the kind of interaction media features that teachers value the most. 
It was filled by teachers that were also students in a Professional Development course (on the use of new technologies in higher education, around 1/4 of them) as well as University of Granada teachers who knew about it by emails from the authors, or from the vicedeanship for International Relations that included it in its newsletter. This means that there might be a higher proportion of 1) teachers with few years of experience 2) teachers from the computer science faculty and 3) teachers from Granada. We think, however that there's no explicit bias in this selection, although of course specific percentages will vary.

Specifically, this survey asked the teachers, in addition to general information (such as gender, age, experience, etc), what are  their \textcolor{black}{past experiences and needs} with regard to the use of instant messaging applications with students, and even using bots or chatbots in the classroom. In addition the survey asked about the most relevant \textit{interaction media features} for the teachers with respect to the potential tool to use in class. 

The specific questions included in this second survey are listed below:


\textit{1. You prefer the interaction with your students to be ...}
\begin{itemize}
\item Synchronous: students and teacher/s in the same space, simultaneously. [\texttt{Synchrony - Yes}]
\item Asynchronous: the student asks at anytime, the teacher responds whenever possible. [\texttt{Synchrony - No}]
\end{itemize}

\textit{2. Regarding the interaction with your students, you ...}
\begin{itemize}
\item Prefer it to be done by institutional means/tools: office hours, forum/messages via the virtual campus, institutional email. [\texttt{Officiality - Yes}]
\item Admit the use of other means/tools: blogs, chat groups in messaging platforms, any others. [\texttt{Officiality - No}]
\end{itemize}

\textit{3. Which media features do you value the most for your interaction with your students?}
\begin{itemize}
\item Familiarity with its use i.e. that I do not need to learn a new tool. [\texttt{Familiarity}]
\item The official formation/training provided by your institution on its use. [\texttt{Official formation}]
\item That only one-directional communication is allowed (from teachers to students). [\texttt{Unidirectionality}]
\item Non-hierarchical tools that allowed teacher-to-student and student-to-student communication. [\texttt{Horizontally}]
\item Connectivity to other tools, for instance common login. [\texttt{Connectivity}]
\item Possibility for teachers to develop/add their own functionalities (e.g. automatic correction tools). [\texttt{Pluggability}]
\item Possibility for users to hide personal data (such as my telephone number). [\texttt{Hidden Phone Number}]
\item Not having to worry about maintenance and that it is easily sustainable from one year to the next. [\texttt{Sustainability}]
\item That it offers the chance to extract data to assess the student performance. [\texttt{Analytics}]
\end{itemize}

\end{backmatter}

\end{document}